\documentstyle[12pt,epsf]{article}
\begin{document}
%\tolerance=5000
%\def\be{\begin{equation}}
%\def\ee{\end{equation}}
%\def\bea{\begin{eqnarray}}
%\def\eea{\end{eqnarray}}
%\def\nn{\nonumber \\}
%\def\det{{\rm det\,}}
%\def\Tr{{\rm Tr\,}}
%\def\e{{\rm e}}
%\def\etal{{\it et al.}}
% Additional Def  
%\newcommand {\be}{\begin{equation}}
\newcommand {\ee}{\end{equation}}
\newcommand {\bea}{\begin{eqnarray}}
\newcommand {\eea}{\end{eqnarray}}
\newcommand {\nn}{\nonumber \\}
\newcommand {\Tr}{{\rm Tr\,}}
\newcommand {\tr}{{\rm tr\,}}
\newcommand {\e}{{\rm e}}
\newcommand {\etal}{{\it et al.}}
% Definition by S.I. 97.6.21, GEGIN
%%%%%%   Abbreviation  %%%%%
\newcommand {\m}{\mu}
\newcommand {\n}{\nu}
\newcommand {\pl}{\partial}
\newcommand {\p} {\phi}
\newcommand {\vp}{\varphi}
\newcommand {\vpc}{\varphi_c}
\newcommand {\al}{\alpha}
\newcommand {\be}{\beta}
\newcommand {\ga}{\gamma}
\newcommand {\Ga}{\Gamma}
\newcommand {\x}{\xi}
\newcommand {\ka}{\kappa}
\newcommand {\la}{\lambda}
\newcommand {\La}{\Lambda}
\newcommand {\si}{\sigma}
\newcommand {\Si}{\Sigma}
\newcommand {\th}{\theta}
\newcommand {\Th}{\Theta}
\newcommand {\om}{\omega}
\newcommand {\Om}{\Omega}
\newcommand {\ep}{\epsilon}
\newcommand {\vep}{\varepsilon}
\newcommand {\na}{\nabla}
\newcommand {\del}  {\delta}
\newcommand {\Del}  {\Delta}
%%%%%%%%%%%
\newcommand {\mn}{{\mu\nu}}
\newcommand {\ls}   {{\lambda\sigma}}
\newcommand {\ab}   {{\alpha\beta}}
\newcommand {\gd}   {{\gamma\delta}}
\newcommand {\half}{ {\frac{1}{2}} }
\newcommand {\third}{ {\frac{1}{3}} }
\newcommand {\fourth} {\frac{1}{4} }
\newcommand {\sixth} {\frac{1}{6} }
\newcommand {\sqg} {\sqrt{g}}
\newcommand {\fg}  {\sqrt[4]{g}}
\newcommand {\invfg}  {\frac{1}{\sqrt[4]{g}}}
\newcommand {\sqZ} {\sqrt{Z}}
\newcommand {\gbar}{\bar{g}}
\newcommand {\sqk} {\sqrt{\kappa}}
\newcommand {\sqt} {\sqrt{t}}
\newcommand {\reg} {\frac{1}{\epsilon}}
\newcommand {\fpisq} {(4\pi)^2}
%%%  cal  %%%%
\newcommand {\Lcal}{{\cal L}}
\newcommand {\Ocal}{{\cal O}}
\newcommand {\Dcal}{{\cal D}}
\newcommand {\Ncal}{{\cal N}}
\newcommand {\Mcal}{{\cal M}}
\newcommand {\scal}{{\cal s}}
%%% vec %%%%
%\newcommand {\Dvec}{{\vec D}}   
\newcommand {\Dvec}{{\hat D}}   %advice by Mike Creutz 98.11.4
\newcommand {\dvec}{{\vec d}}
\newcommand {\Evec}{{\vec E}}
\newcommand {\Hvec}{{\vec H}}
\newcommand {\Vvec}{{\vec V}}
%%%%  til  %%%%
\newcommand {\Btil}{{\tilde B}}
\newcommand {\ctil}{{\tilde c}}
\newcommand {\Ftil}{{\tilde F}}
\newcommand {\Ktil}  {{\tilde K}}
\newcommand {\Ltil}  {{\tilde L}}
\newcommand {\mtil}{{\tilde m}}
\newcommand {\ttil} {{\tilde t}}
\newcommand {\Qtil}  {{\tilde Q}}
\newcommand {\Rtil}  {{\tilde R}}
\newcommand {\Stil}{{\tilde S}}
\newcommand {\Ztil}{{\tilde Z}}
\newcommand {\altil}{{\tilde \alpha}}
\newcommand {\betil}{{\tilde \beta}}
\newcommand {\etatil} {{\tilde \eta}}
\newcommand {\latil}{{\tilde \lambda}}
\newcommand {\ptil}{{\tilde \phi}}
\newcommand {\Ptil}{{\tilde \Phi}}
\newcommand {\natil} {{\tilde \nabla}}
\newcommand {\xitil} {{\tilde \xi}}
%%%%  hat   %%%%%%
\newcommand {\Rhat}{{\hat R}}
\newcommand {\Shat}{{\hat S}}
\newcommand {\ehat}{{\hat e}}
\newcommand {\mhat}{{\hat m}}
\newcommand {\shat}{{\hat s}}
\newcommand {\Dhat}{{\hat D}}   
\newcommand {\Vhat}{{\hat V}}   
\newcommand {\xhat}{{\hat x}}
\newcommand {\Zhat}{{\hat Z}}
\newcommand {\Gahat}{{\hat \Gamma}}
\newcommand {\nah} {{\hat \nabla}}
\newcommand {\etahat} {{\hat \eta}}
\newcommand {\omhat} {{\hat \omega}}
\newcommand {\psihat} {{\hat \psi}}
\newcommand {\thhat} {{\hat \theta}}
\newcommand {\gh}  {{\hat g}}
%%%%  bar  %%%%
\newcommand {\Kbar}  {{\bar K}}
\newcommand {\Lbar}  {{\bar L}}
\newcommand {\Qbar}  {{\bar Q}}
\newcommand {\labar}{{\bar \lambda}}
\newcommand {\cbar}{{\bar c}}
\newcommand {\bbar}{{\bar b}}
\newcommand {\Bbar}{{\bar B}}
\newcommand {\psibar}{{\bar \psi}}
\newcommand {\chibar}{{\bar \chi}}
\newcommand {\fbar}{{\bar 5}}
\newcommand {\bbartil}{{\tilde {\bar b}}}
%%%%%%%%%  number suffix
\newcommand  {\vz}{{v_0}}
\newcommand  {\ez}{{e_0}}
\newcommand  {\mz}{{m_0}}
%%%%  Integral  %%%%%%%%%%%
\newcommand {\intfx} {{\int d^4x}}
\newcommand {\inttx} {{\int d^2x}}
%%%%%%%%%%%%%%%%%%%%%%%%%%%%%%%%%%%%%%%%%%%%%%%%%%
\newcommand {\change} {\leftrightarrow}
\newcommand {\ra} {\rightarrow}
\newcommand {\larrow} {\leftarrow}
\newcommand {\ul}   {\underline}
\newcommand {\pr}   {{\quad .}}
\newcommand {\com}  {{\quad ,}}
\newcommand {\q}    {\quad}
\newcommand {\qq}   {\quad\quad}
\newcommand {\qqq}   {\quad\quad\quad}
\newcommand {\qqqq}   {\quad\quad\quad\quad}
\newcommand {\qqqqq}   {\quad\quad\quad\quad\quad}
\newcommand {\qqqqqq}   {\quad\quad\quad\quad\quad\quad}
\newcommand {\qqqqqqq}   {\quad\quad\quad\quad\quad\quad\quad}
\newcommand {\lb}    {\linebreak}
\newcommand {\nl}    {\newline}

%%%%%%%%%%% Space  %%%%%%%
\newcommand {\vs}[1]  { \vspace*{#1 cm} }

%%%%%%%%%%%%   Journal %%%%%%%%%%%%%%
\newcommand {\MPL}  {Mod.Phys.Lett.}
\newcommand {\NP}   {Nucl.Phys.}
\newcommand {\PL}   {Phys.Lett.}
\newcommand {\PR}   {Phys.Rev.}
\newcommand {\PRL}   {Phys.Rev.Lett.}
\newcommand {\CMP}  {Commun.Math.Phys.}
\newcommand {\JMP}  {Jour.Math.Phys.}
\newcommand {\AP}   {Ann.of Phys.}
\newcommand {\PTP}  {Prog.Theor.Phys.}
\newcommand {\NC}   {Nuovo Cim.}
\newcommand {\CQG}  {Class.Quantum.Grav.}

% Definition by S.I. 97.6.21,  END

\font\smallr=cmr5
%%%%%%%%%%%%%%%%%%%% definition  by IKEDA  ,SEC 4.5  %%%%%%%%%%%%%%%%%%%%%%%
\def\ocirc#1{#1^{^{{\hbox{\smallr\llap{o}}}}}}
\def\ogamma{\ocirc{\gamma}{}}
\def\oM{{\buildrel {\hbox{\smallr{o}}} \over M}}
\def\osigma{\ocirc{\sigma}{}}

\def\overleftrightarrow#1{\vbox{\ialign{##\crcr
 $\leftrightarrow$\crcr\noalign{\kern-1pt\nointerlineskip}
 $\hfil\displaystyle{#1}\hfil$\crcr}}}
\def\overnab{{\overleftrightarrow\nabslash}}

\def\va{{a}}
\def\vb{{b}}
\def\vc{{c}}
\def\tilpsi{{\tilde\psi}}
\def\tbpsi{{\tilde{\bar\psi}}}
%%%%%%%%%%%%%%%%%%%% definition  by IKEDA  ,SEC 4.5, App.A  %%%%%%%%%%%%%%%%%%%
%%%%%%%%%%%%%%%%%%%%%%%%%%%%%%%%%%%%%%%%%%%%%%%%%%%%%%%%%%%%%%%%%%%%%%%%%%%%%%

\def\delL{{\delta_{LL}}}
\def\delG{{\delta_{G}}}
\def\delc{{\delta_{cov}}}

%%%%%%%%%%%%%%%%%%%% definition  by IKEDA , SEC 5  %%%%%%%%%%%%%%%%%%%%%%%
%%%%%%%%%%%%%%%%%%%% 98.10.30 Chiral Fermion, with Creutz  %%%%%%%%%%%%%%%%
\newcommand {\sqxx}  {\sqrt {x^2+1}}   %99.2.20 wall.tex App.B
\newcommand {\gago}  {\gamma^5}
\newcommand {\Pp}  {P_+}
\newcommand {\Pm}  {P_-}
\newcommand {\GfMp}  {G^{5M}_+}
\newcommand {\GfMpm}  {G^{5M'}_-}
\newcommand {\GfMm}  {G^{5M}_-}
\newcommand {\Omp}  {\Omega_+}    %99.4.17 wall.tex
\newcommand {\Omm}  {\Omega_-}
\def\Aslash{{}\hbox{\hskip2pt\vtop
 {\baselineskip23pt\hbox{}\vskip-24pt\hbox{/}}
 \hskip-11.5pt $A$}}
\def\Rslash{{}\hbox{\hskip2pt\vtop
 {\baselineskip23pt\hbox{}\vskip-24pt\hbox{/}}
 \hskip-11.5pt $R$}}
\def\kslash{
{}\hbox       {\hskip2pt\vtop
                   {\baselineskip23pt\hbox{}\vskip-24pt\hbox{/}}
               \hskip-8.5pt $k$}
           }    %99.2.23 -11.5 -> -8.5
\def\qslash{
{}\hbox       {\hskip2pt\vtop
                   {\baselineskip23pt\hbox{}\vskip-24pt\hbox{/}}
               \hskip-8.5pt $q$}
           }    
\def\dslash{
{}\hbox       {\hskip2pt\vtop
                   {\baselineskip23pt\hbox{}\vskip-24pt\hbox{/}}
               \hskip-8.5pt $\partial$}
           }    
\def\dbslash{{}\hbox{\hskip2pt\vtop
 {\baselineskip23pt\hbox{}\vskip-24pt\hbox{$\backslash$}}
 \hskip-11.5pt $\partial$}}
\def\Kbslash{{}\hbox{\hskip2pt\vtop
 {\baselineskip23pt\hbox{}\vskip-24pt\hbox{$\backslash$}}
 \hskip-11.5pt $K$}}
\def\Ktilbslash{{}\hbox{\hskip2pt\vtop
 {\baselineskip23pt\hbox{}\vskip-24pt\hbox{$\backslash$}}
 \hskip-11.5pt ${\tilde K}$}}
\def\Ltilbslash{{}\hbox{\hskip2pt\vtop
 {\baselineskip23pt\hbox{}\vskip-24pt\hbox{$\backslash$}}
 \hskip-11.5pt ${\tilde L}$}}
\def\Qtilbslash{{}\hbox{\hskip2pt\vtop
 {\baselineskip23pt\hbox{}\vskip-24pt\hbox{$\backslash$}}
 \hskip-11.5pt ${\tilde Q}$}}
\def\Rtilbslash{{}\hbox{\hskip2pt\vtop
 {\baselineskip23pt\hbox{}\vskip-24pt\hbox{$\backslash$}}
 \hskip-11.5pt ${\tilde R}$}}
\def\Kbarbslash{{}\hbox{\hskip2pt\vtop
 {\baselineskip23pt\hbox{}\vskip-24pt\hbox{$\backslash$}}
 \hskip-11.5pt ${\bar K}$}}
\def\Lbarbslash{{}\hbox{\hskip2pt\vtop
 {\baselineskip23pt\hbox{}\vskip-24pt\hbox{$\backslash$}}
 \hskip-11.5pt ${\bar L}$}}
\def\Rbarbslash{{}\hbox{\hskip2pt\vtop
 {\baselineskip23pt\hbox{}\vskip-24pt\hbox{$\backslash$}}
 \hskip-11.5pt ${\bar R}$}}
\def\Qbarbslash{{}\hbox{\hskip2pt\vtop
 {\baselineskip23pt\hbox{}\vskip-24pt\hbox{$\backslash$}}
 \hskip-11.5pt ${\bar Q}$}}
\def\Acalbslash{{}\hbox{\hskip2pt\vtop
 {\baselineskip23pt\hbox{}\vskip-24pt\hbox{$\backslash$}}
 \hskip-11.5pt ${\cal A}$}}

\begin{flushright}
June 2002\\
hep-th/0206187 \\
US-02-05
\end{flushright}

\vspace{0.5cm}

\begin{center}

{\Large\bf 
Fermions in\\
Kaluza-Klein and Randall-Sundrum Theories}\footnote
{This work is based on the content presented at
the Fifth KEK Topical Conference (2001.11.19-22, Tsukuba, Japan) 
\cite{KEK01}
}

\vspace{1.5cm}
%{\large Note by S.I.}
{\large Shoichi ICHINOSE
         \footnote{
E-mail address:\ ichinose@u-shizuoka-ken.ac.jp
                  }
}
\vspace{1cm}

{\large 
Laboratory of Physics, \\
School of Food and Nutritional Sciences, \\
University of Shizuoka,
Yada 52-1, Shizuoka 422-8526, Japan          }

\end{center}

\vfill

{\large Abstract}\nl
The Kaluza-Klein theory and Randall-Sundrum
theory are examined comparatively, with focus
on the behavior of the five dimensional (Dirac) fermion
in the dimensional reduction to four dimensions.
They are properly treated using the Cartan formalism.
In the KK case, the dual property between the electric and magnetic
dipole moments is revealed in relation to the ratio of
two massive parameters:\ the inverse of the radius of the extra-space
circle and  the 5D fermion mass. The order estimation
of the couplings is done.
In the RS case, we consider the interaction with the 5D(bulk) Higgs field
and the gauge field.  
The chiral property, localization, anomaly phenomena
are examined. 
We evaluate the bulk quantum effect using the method of
the induced effective action. The electric dipole
moment term naturally appears.  This is a new
origin of the CP-violation. In the 4D limit, 
the dual relation between KK model
and RS model appears. 

\vspace{0.5cm}

PACS NO:
04.50.+h,\ 
%Gravity in more than four dimensions, Kaluza-Klein theory, 
%unified field theories, alternative theories of
11.10.Kk,\ 
%Field theories in dimensions other than four 
11.25.Mj,\ 
%Compactification and four-dimensional models
12.10.-g 
%Unified field theories and models 
11.30.Er,\ 
%Charge conjugation,parity,time reversal, and other discrete
%symmetries
13.40.Em,\ 
%Electric and magnetic moments
\nl

%\newpage

%%%%%%%%%%%%%%%%%%%%%%%%%%%%%%%%%%%%%%%%%%%%%%%%%%%%%%%%%%%%%%%%%%%%%%
%%%%%%%%%%%%%%%%%%%%%%%  Sec.1  Intro  %%%%%%%%%%%%%%%%%%%%%%%%%%%%%%%
%%%%%%%%%%%%%%%%%%%%%%%%%%%%%%%%%%%%%%%%%%%%%%%%%%%%%%%%%%%%%%%%%%%%%%
\section{Introduction}
If the present research direction of the unification
of forces, using strings and D-branes, is right, 
the unified theory should be, effectively at some scale, 
some higher-dimensional field theory. Then 
the real world of 4 dimensions could be viewed as some ``approximation"
of the higher dimensional one. 
This procedure is called {\it dimensional reduction}. 
There are two 
representative and contrastive
approaches:\ Kaluza-Klein (KK) and Randall-Sundrum (RS)
theories. At present 
there seems to be no strong reason
to believe that one is better than the other. 
Both have advantages and disadvantages.
From the viewpoint of the space-time unification of gauge theories, 
KK reduction looks attractive. 
The gauge fields are nicely realized from the space-time symmetry.
On the other hand, RS reduction looks attractive 
in some physically-interesting points such as
the chirality control, the massless localized fermion and the anomaly flow.
In fact many literatures on both theories
have been produced and are now being produced. 
In the present paper we compare the two approaches
and reveal their important features by contrasting them.

The Kaluza-Klein theory has the long history.\cite{Kal21, Klein26} 
It is characterized by {\it compactifying}
the extra manifold. In this procedure 
the radius of the compact manifold, $1/\mu$, is introduced
as the size parameter. 
On the other hand, in the Randall-Sundrum model\cite{RS9905,RS9906},
the {\it localized} configuration in the extra space is utilized, 
instead of compactifying the extra space. In this procedure
the size parameter, $1/k$, of the localization 
("thickness" of the wall) is introduced. 
Both approaches accomplish the dimensional reduction
by adjusting the size parameters.

In the original papers by Randall and Sundrum, 
the comparison between the two models is made mainly in the
mass hierarchy. 
Here we focus on another aspect, that is, 
the {\it magnetic and electric dipole moment} terms. 
They are both higher-dimensional operators (the mass-dimension 5). 
The former one usually appears, in the 4D QED,
as a {\it quantum} effect. 
The latter one is a {\it CP-violation} term. Both ones
are very important in phenomenology and experiments.
(Their measurement could become a strong candidate to
confirm a higher dimensional unification model.
Estimation of the dipole moments for various
elementary particles, from the naturalness requirement, 
has been done in a recent interesting paper \cite{AHK0111}.)
The appearance of the CP-violating term in the KK model
was presented by Thirring very long ago\cite{Thirr72}.
The results are still important and interesting  
even at the present time, but the work was done
so many years ago that the content is not familiar
to most readers. Therefore we explain it, 
in the present viewpoint, in Sec.2 and 3. 
We regard it as a proto-type to investigate the dimensional
reduction and use it  
for the comparison with the RS case.
We analyze the higher-dimensional terms in the RS model. 
The estimated magnitude of the electric dipole moment (EDM)  
and the magnetic dipole moment (MDM) in KK model is too small
to observe, while that in the RS model can be expected to
escape the case. As other approaches of
CP-violation, in the brane world context, we refer to  
ref.\cite{Saka0011,DEK0202}.

In order to properly treat fermions in the general relativity, 
we take the {\it Cartan formalism}\cite{Car28}
(See a textbook\cite{MTW73, Wald84, Naka91}. %Eguchi-Gilkey-Hanson?
). 
It clarifies
the {\it gauge structure} of the system on the local Lorentz frame.
We explain it both for the KK and for the RS geometries. 
Recently many brane world models are discussed, where the
fermion part is important in relation to the 
{\it chiral property}. 
%Some of them, however, are dubious in or hide
%the systematic treatment of the fermion part.
Because the chiral property much depends on the delicate
points such as the signs ($\pm$) or the phases ($\pm i$),
it is worthwhile presenting a strict formulation of the fermion theory
at the present stage of the development.
In the analysis, we take a {\it stable} brane world model\cite{RS9906, SI0003},
that is {\it one} wall RS model.

The paper consists of the following contents. In Sec.2 we review
the KK theory in the Cartan formalism. 
It is compared with the case of the RS theory in Sec.4 and App.B.
Sec.3 treats the Dirac fermion in KK theory, where
the electric and magnetic dipole moments appear and 
the {\it dual} property is stressed.
In Sec.4, Randall-Sundrum theory is explained in the Cartan
formalism. 
The Dirac fermion in the RS theory is explained in Sec.5.
The RS counterpart of the KK expansion appears. 
The eigen functions are characterized by the Bessel differential
equation. 
The bulk (5D) Higgs mechanism is explained in Sec.6, where
the {\it localization} of the fermion is explained.
The fermion gets a mass through the Yukawa interaction with
the bulk Higgs. 
The 5D QED is examined in Sec.7, where 
the 5D {\it bulk quantum} effects produce the {\it anomaly flow}
and the {\it electric dipole moment} in the effective action. 
We conclude in Sec.8.
Two appendixes are provided for clarifying the text.

%**************************************************

%discrete symmetry

%The origin of the CP-violation has been discussed
%so long time. Thirring 72,'tHooft 76, Weinberg ?,
%We examine the aspect from the higher dimensional
%models. 

%Appelquist-Chodos, Vaccum energy ?
%Recent a few years, triggered by Randall and Sundrum, 
%there has been stressed the domain world(DW) approach. 

%%%%%%%%%%%%%%%%%%%%%%%%%%%%%%%%%%%%%%%%%%%%%%%%%%%%%%%%%%%%%%%%%%%%%%
%%%%%%%%%%%%%%%%%%%%%%    SEC. 2  Kaluza-Klein Theory     %%%%%%%%%%%
%%%%%%%%%%%%%%%%%%%%%%%%%%%%%%%%%%%%%%%%%%%%%%%%%%%%%%%%%%%%%%%%%%%%%%
\section{Kaluza-Klein Theory}

Let us first review the 5D Kaluza-Klein theory. 
This serves as the preparation for a similar treatment of 
the Randall-Sundrum theory.
The 5D space-time manifold is described by the 4D coordinates
$x^a$ ($a=0,1,2,3$) and an {\it extra} coordinate $y$. We also use
the notation ($X^m$)=($x^a, y$), ($m=0,1,2,3,5$). 
With the general 5D metric $\gh_{mn}$,
%*** KK1%%%%%%%%%%%%%%%%%%%%
\begin{eqnarray}
ds^2=\gh_{mn}(X)dX^mdX^n
\com
\label{KK1}
\end{eqnarray}
%%%%%%%%%%%%%%%%%%%%%%%%%%%%%
we assume the $S^1$ compactification condition for the
extra space.
%*** KK1b%%%%%%%%%%%%%%%%%%%%
\begin{eqnarray}
\gh_{mn}(x,y)=\gh_{mn}(x,y+\frac{2\pi}{\mu})
\com
\label{KK1b}
\end{eqnarray}
%%%%%%%%%%%%%%%%%%%%%%%%%%%%%
where $\mu^{-1}$ is the {\it radius} of the extra space circle.
We specify the form of the metric as
%*** KK1c%%%%%%%%%%%%%%%%%%%%
\begin{eqnarray}
ds^2=
g_{ab}(x)dx^adx^b+\e^{2\si(x)}(dy-fA_a(x)dx^a)^2
\com
\label{KK1c}
\end{eqnarray}
%%%%%%%%%%%%%%%%%%%%%%%%%%%%%
where $g_{ab}(x), A_a(x)$ and $\si(x)$ are all 4D quantities, namely,  
the 4D metric, the U(1) gauge field and the dilaton (Weyl scale) field, 
respectively. 
$f$ is a coupling constant. 
This specification is based on the
following additional assumptions.

\begin{description}
\item[1.]
$y$ is a space coordinate.
\item[2.]
The geometry is invariant under
the U(1) symmetry:\ 
$y\ra y+\La(x),\ A_a(x)\ra A_a(x)+\frac{1}{f}\pl_a\La$ .
%\item[3.]
%When the gauge field $A_a(x)$ switches off ( or $f=0$),
%the 4D world \{$x^a$\} and the extra one \{$y$\} are
%decomposable (separate worlds). 
\end{description}
The $S^1$ extra space \{$y$\} is here identified as the 
"U(1) gauge parameter space". 
(This way of unification of the gauge theory and gravity 
should be compared with the treatment in the RS approach
where the gauge parameter space is introduced as an {\it internal} one.
)

We can generalize (\ref{KK1c}), 
keeping the periodicity condition (\ref{KK1b}), 
by taking the following
ones instead of $g_{ab}(x), A_a(x)$ and $\si(x)$
of (\ref{KK1c}),
%*** KK2 %%%%%%%%%%%%%%%%%%%%
\begin{eqnarray}
g_{ab}(X)=\sum_{k\in {\bf Z}}{\tilde g}^{(k)}_{ab}(x)\e^{ik\m y},
A_{a}(X)=\sum_{k\in {\bf Z}}{\tilde A}^{(k)}_{a}(x)\e^{ik\m y},
\si(X)=\sum_{k\in {\bf Z}}{\tilde \si}^{(k)}(x)\e^{ik\m y},
\label{KK2}
\end{eqnarray}
%%%%%%%%%%%%%%%%%%%%%%%%%%%%%
where ${\bf Z}$ is the set of all integers.
The appearance of the {\it periodic} function $\{\e^{ik\m y}|k\in {\bf Z}\}$
is a character of the KK approach. 
The eq.(\ref{KK1c}) is the zero-th ($k=0$, massless) mode of (\ref{KK2}).
%The expression (\ref{KK1c}) is the massless mode (k=0) of (\ref{KK2}). 
We do not examine, in this paper,
the role of the {\it massive} modes (k$\neq$0) for these boson
fields. 

We take the Cartan formalism to introduce Dirac fermions and 
to compute the geometric
quantities such as the connection and 
the Riemann curvature\cite{Thirr72}. 
The basis \{$\thhat^\m$\} of the cotangent manifold($T_p^*M$)
can be defined as the 1-form which satisfies
the relation:
%*** KK3 %%%%%%%%%%%%%%%%%%%%
\begin{eqnarray}
ds^2=\thhat^\m \thhat^\n \etahat_\mn \com\q
\mbox{diag}(\etahat_\mn)=(-1,1,1,1,1),
\label{KK3}
\end{eqnarray}
%%%%%%%%%%%%%%%%%%%%%%%%%%%%%
where $\m,\n= 0,1,2,3,\fbar $ are the local Lorentz (tangent frame)
indices. 
$ds^2$ is given in (\ref{KK1c}). 
The f\"{u}nf-bein $\ehat^\m_{~m}$ is generally defined as
%*** KK4 %%%%%%%%%%%%%%%%%%%%
\begin{eqnarray}
\thhat^\m=\ehat^\m_{~m}dX^m
\pr
\label{KK4}
\end{eqnarray}
%%%%%%%%%%%%%%%%%%%%%%%%%%%%%
$\{\thhat^\m\}$ is the dual of 
the basis $\{\ehat^\m\equiv\ehat^{\m m}\pl/\pl X^m\}$
of the tangent manifold ($T_pM$).
($<\ehat^\m,\thhat_\n>=\del^\m_\n$.) 
For the 4D part, we denote as $\thhat^\al=\theta^\al$
($\al=0,1,2,3$)
which satisfies
%*** KK5 %%%%%%%%%%%%%%%%%%%%
\begin{eqnarray}
g_{ab}(x)dx^adx^b=e^\al_{~a}(x) e^\be_{~b}(x)\eta_\ab dx^adx^b
=\theta^\al\theta^\be\eta_\ab\com\nn
\theta^\al=e^\al_{~a}(x)dx^a
\com
\label{KK5}
\end{eqnarray}
%%%%%%%%%%%%%%%%%%%%%%%%%%%%%
where $g_{ab}$ is the same as (\ref{KK1c}), 
$e^\al_{~a}$ and $\eta_\ab$ are
the 4D part of the f\"{u}nf-bein $\ehat^\m_{~m}$ and 
the 5D flat (Minkowski) metric $\etahat_\mn$ respectively.
($e^\al_{~a}=\ehat^\al_{~a}, \eta_\ab=\etahat_\ab$)
\footnote{
The suffix notation taken in this article is as follows:\ 
$m,n,\cdots$ (Roman alphabets beginning from the middle) are for
1+4 D general space-time coordinates;\ 
$a,b,\cdots$ (Roman alphabets beginning from the first) are for
1+3 D general space-time coordinates;\ 
$\m,\n,\cdots$ (Greek symbols beginning from the middle) are for
1+4 D local Lorentz (Minkowski) coordinates;\ 
$\al,\be,\cdots$ (Greek symbols beginning from the first) are for
1+3 D local Lorentz (Minkowski) coordinates.
The 5-th component is explicitly denoted as "$5$" for the
general coordinate and as $\fbar$ for the local
Lorentz one. The hat symbol 
"$~\hat{\mbox{ }}~$" 
is used to discriminate 5D symbols
from 4D symbols when it is necessary.}
The extra component $\thhat^\fbar$ can be read off, from (\ref{KK1c}), as
%*** KK6 %%%%%%%%%%%%%%%%%%%%
\begin{eqnarray}
\thhat^\fbar=e^\si (dy-fA_adx^a)  \pr
\label{KK6}
\end{eqnarray}
%%%%%%%%%%%%%%%%%%%%%%%%%%%%%
Hence the f\"{u}nf-bein $\ehat^\m_{~m}$ 
and its inverse $\ehat_\m^{~m}$ can be concretely given by
%*** KK7 %%%%%%%%%%%%%%%%%%%%
\begin{eqnarray}
(\ehat^\m_{~m})=\left(
\begin{array}{cc}
e^\al_{~a} & 0 \\
-fe^\si A_a   & e^\si
\end{array}
                 \right),\ 
(\ehat_\m^{~m})=\left(
\begin{array}{cc}
e_\al^{~a} & 0 \\
fA_\al   & e^{-\si}
\end{array}
                 \right),\nn
\ehat^\m_{~m}\ehat_\m^{~n}=\del^n_m\com\q
\ehat^\m_{~m}\ehat_\n^{~m}=\del^\m_\n\com\q
A_\al\equiv e_\al^{~a}A_a
                 \pr
\label{KK7}
\end{eqnarray}
%%%%%%%%%%%%%%%%%%%%%%%%%%%%%

The first Cartan's structure equation, for the torsionless
case(${\hat T}^\m=0$), is given by
%*** KK8 %%%%%%%%%%%%%%%%%%%%
\begin{eqnarray}
d\thhat^\m+\omhat^\m_{~\n}\wedge\thhat^\n={\hat T}^\m=0
\com\q
\omhat_\mn+\omhat_{\n\m}=0
\com
\label{KK8}
\end{eqnarray}
%%%%%%%%%%%%%%%%%%%%%%%%%%%%%
where $\omhat^\m_{~\n}$ is the {\it connection 1-form}. 
From this equation and eqs.(\ref{KK5}) and (\ref{KK6}), 
we can read the connection as
%*** KK9 %%%%%%%%%%%%%%%%%%%%
\begin{eqnarray}
\omhat^\fbar_{~\fbar}%=-\half\pl_a\si e_\al^{~a}\thhat^\al
=-\half\pl_a\si~dx^a
\com\q 
\omhat^\fbar_{~\al}=-\omhat_\al^{~\fbar}
=\half\pl_a\si~e_\al^{~a}\thhat^\fbar -\frac{f}{2}e^\si F_\ab\thhat^\be\com\nn
\omhat^\al_{~\be}=\om^\al_{~\be}
+\frac{f}{2}e^\si F^\al_{~\be}~\thhat^\fbar\com\nn
F_{ab}=\pl_aA_b-\pl_bA_a\com\q
F_\ab\equiv e_\al^{~a}e_\be^{~b}F_{ab}
\com
\label{KK9}
\end{eqnarray}
%%%%%%%%%%%%%%%%%%%%%%%%%%%%%
where $\om^\al_{~\be}$ is the 4D connection which is
defined by $d\theta^\al+\om^\al_{~\be}\wedge\th^\be=0$.
(Note that $\om^\al_{~\be}\neq \omhat^\al_{~\be}$.)

The second Cartan's structure equation is given by
%*** KK10 %%%%%%%%%%%%%%%%%%%%
\begin{eqnarray}
d\omhat_\mn+\omhat_{\m\si}\wedge\omhat^\si_{~\n}=
\half\Rhat_{\mn\si\tau}\thhat^\si\wedge\thhat^\tau
\pr
\label{KK10}
\end{eqnarray}
%%%%%%%%%%%%%%%%%%%%%%%%%%%%%
This gives the curvatures $\Rhat_{\mn\si\tau}$ as
%*** KK11 %%%%%%%%%%%%%%%%%%%%
\begin{eqnarray}
\Rhat_{\ab\ga\del}=R_{\ab\ga\del}-\frac{f^2}{2}e^{2\si}F_\ab F_{\ga\del}
+\frac{f^2}{4}e^{2\si}(F_{\al\del}F_{\be\ga}-\del\change\ga)\ ,\nn
\Rhat_{\al \fbar\fbar\be}=\frac{f^2}{4}e^{2\si}F_{\al\ga}F^\ga_{~\be}
+\cdots \pr
\label{KK11}
\end{eqnarray}
%%%%%%%%%%%%%%%%%%%%%%%%%%%%%
Hence the 5D Riemann scalar curvature
$\Rhat^{\m~\n}_{~\n~\m}$ can be decomposed as
%*** KK12 %%%%%%%%%%%%%%%%%%%%
\begin{eqnarray}
\Rhat=R+\frac{f^2}{4}e^{2\si}F^\ab F_\ab
+\half\pl_a\si\pl_b\si g^{ab}+\half\na^2\si
\com
\label{KK12}
\end{eqnarray}
%%%%%%%%%%%%%%%%%%%%%%%%%%%%%
which shows the theory of {\it gravity}, 
{\it electro-magnetism} and the dilaton in 
the 4D world. The correct sign appears in front of 
$F^\ab F_\ab$, which comes from the choice of the extra coordinate
$y$, in (\ref{KK1c}), as a {\it space} (not time) component.
This point was stressed by Thirring\cite{Thirr72}.

We consider the simple case $\si=0$ for the present purpose. 
%%%%%%%%%%%%%%%%%%%%%%%%%%%%%%%%%%%%%%%%%%%%%%%%%%%%%%%%%%%%%%%%%%%%%
%%%% SEC. 3  Fermions in Kaluza-Klein Theory  %%%%%%%%%%%%%%%%%%%%%%%
%%%%%%%%%%%%%%%%%%%%%%%%%%%%%%%%%%%%%%%%%%%%%%%%%%%%%%%%%%%%%%%%%%%%%
\section{Fermions in Kaluza-Klein Theory}
The 5D Dirac equation is generally given by
(see a textbook\cite{MTW73, Wald84, Naka91})
%*** ferKK1 %%%%%%%%%%%%%%%%%%%%
\begin{eqnarray}
\left\{
\ga^\m\ehat_\m^{~m}\frac{\pl}{\pl X^m}
+\frac{1}{8}(\omhat^\si)_\mn\ga_\si [\ga^\m,\ga^\n]+\mhat
\right\}\psihat=0
\pr
\label{ferKK1}
\end{eqnarray}
%%%%%%%%%%%%%%%%%%%%%%%%%%%%%
$\mhat$ is the mass parameter of the 5D fermion 
($-\infty<\mhat<\infty$).
\footnote{
The 5D gamma matrices $\{\ga^\m\}$ are defined in App.A.
}
The spin connection above $(\omhat^\si)_\mn$ is defined by
$\omhat^\mu_{~\nu}=(\omhat_\la)^\mu_{~\nu}\thhat^\la$.
(Local Lorentz suffixes are lowered or raised by the
5D Minkowski metric $\etahat_\mn$ or $\etahat^\mn$ respectively.)
The 5D Dirac matrix $\ga^\m$ satisfies 
$\{\ga^\m,\ga^\n\}=2\etahat^\mn$
%, \etahat^\mn=\mbox{diag}(-1,1,1,1,1)$
. 
For simplicity we switch off the 4D gravity: 
$e^\al_{~a}\ra\del^\al_a\ ,\ \om^\al_{~\be}\ra 0$. 
In the present case, using (\ref{KK7}) and (\ref{KK9}),
the eq.(\ref{ferKK1}) says
%*** ferKK2 %%%%%%%%%%%%%%%%%%%%
\begin{eqnarray}
\left\{
\ga^a(\pl_a+fA_a\pl_5)+\ga^5\pl_5
-\frac{f}{16}F_{ab}\ga^5[\ga^a,\ga^b]+\mhat
\right\}\psihat=0
\com
\label{ferKK2}
\end{eqnarray}
%%%%%%%%%%%%%%%%%%%%%%%%%%%%%
where $\pl_5=\pl/\pl y$. \nl
\nl
(A) Charged Fermion\nl
Corresponding to a {\it massive} mode in the {\it 4D reduction}
(\ref{KK2}), we consider the following form for a charged fermion.
%*** ferKK3 %%%%%%%%%%%%%%%%%%%%
\begin{eqnarray}
\psihat(x,y)=\e^{i(\phi\gago+\m y)}\psi(x)
\pr
\label{ferKK3}
\end{eqnarray}
%%%%%%%%%%%%%%%%%%%%%%%%%%%%%
Here we regard the charged fermion as a KK-{\it massive} mode.
The {\it angle parameter} $\phi$ is chosen as
%*** ferKK4 %%%%%%%%%%%%%%%%%%%%
\begin{eqnarray}
(i\gago\m+\mhat)\e^{2i\phi\gago}
=\sqrt{{\mhat}^2+\m^2}\equiv M\pr\nn
(i)\ \mhat\neq 0\ :\q\tan 2\phi=-\frac{\m}{\mhat}\ ,\nn
-\frac{\pi}{2}<2\phi\leq 0 \ \ \mbox{for}\ \mhat>0;\ \ 
\pi\leq 2\phi< \frac{3\pi}{2}\ \ \mbox{for}\ \mhat<0
\pr\nn
(ii)\ \mhat=0\ :\q 2\phi=-\frac{\pi}{2}\com
\label{ferKK4}
\end{eqnarray}
%%%%%%%%%%%%%%%%%%%%%%%%%%%%%
where $(\gago)^2=1, e^{i\al\gago}=\cos\al+i\gago\sin\al,$ is used. 
See the lower half region of Fig.1.
%%%%%%%%%%%%%%%%%  Fig.1 %%%%%%%%%%%%%%%%%%%%%
\begin{figure}%[htb]
\epsfysize=5cm\epsfbox{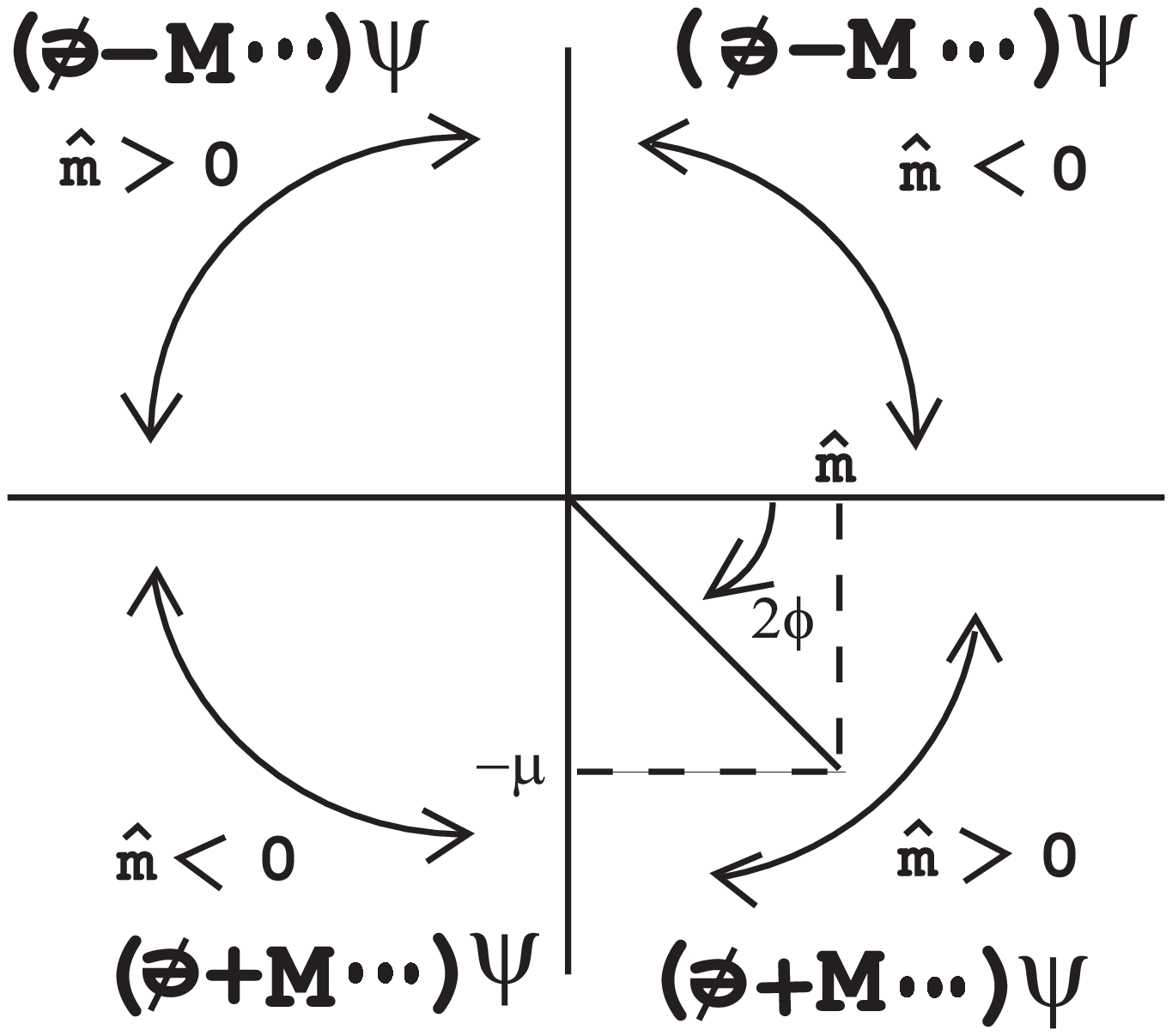}
   \begin{center}
Fig.1\ The angle parameter $\phi$ which defines the 4D charged
fermion in the KK dimensional reduction (\ref{ferKK3}).
$\mhat$ is the 5D fermion mass parameter, $\m^{-1}$ is the
size of the extra compact manifold. 
$-\pi/2 <2\phi\leq 0\ \mbox{for}\ \mhat>0;\  
\pi \leq 2\phi< 3\pi/2\ \mbox{for}\ \mhat<0;\ 
2\phi=-\pi/2\ \mbox{for}\ \mhat=0$.
The upper half region gives the same 4D fermion action
as the lower half region by the transformation $\psi\ra\gago\psi$.
%***angle2.eps
   \end{center}
\label{fig:angle}
\end{figure}
%%%%%%%%%%%%%%%%%%%%%%%%%%%%%%%%%%%%%%%%%%%%%%%

Then (\ref{ferKK2}) reduces to
%*** ferKK5 %%%%%%%%%%%%%%%%%%%%
\begin{eqnarray}
\left\{
\ga^a(\pl_a+ieA_a)+M
-\frac{1}{16M}(\mhat\frac{e}{\m}-ie\gago)F_{ab}\gago [\ga^a,\ga^b]
\right\}\psi=0
\com
\label{ferKK5}
\end{eqnarray}
%%%%%%%%%%%%%%%%%%%%%%%%%%%%%
where $e\equiv f\m$ is the {\it electric coupling constant}.
$M$ is identified as a 4D fermion mass.
\footnote{
We can take the angle $\phi$ in (\ref{ferKK4}) in another way.
%*** ferKK4x %%%%%%%%%%%%%%%%%%%%
\begin{eqnarray}
(i\gago\m+\mhat)\e^{2i\phi\gago}
=-\sqrt{{\mhat}^2+\m^2}=-M\pr\nn
(i)\ \mhat\neq 0\ :\q\tan 2\phi=-\frac{\m}{\mhat}\ ,\nn
\frac{\pi}{2}<2\phi\leq \pi \ \ \mbox{for}\ \mhat>0;\ \ 
0\leq 2\phi< \frac{\pi}{2}\ \ \mbox{for}\ \mhat<0
\pr\nn
(ii)\ \mhat=0\ :\q 2\phi=\frac{\pi}{2}\com
\label{ferKK4x}
\end{eqnarray}
%%%%%%%%%%%%%%%%%%%%%%%%%%%%%
Then the 4D equation (\ref{ferKK5}) is replaced by
%*** ferKK5x %%%%%%%%%%%%%%%%%%%%
\begin{eqnarray}
\left\{
\ga^a(\pl_a+ieA_a)-M
+\frac{1}{16M}(\mhat\frac{e}{\m}-ie\gago)F_{ab}\gago [\ga^a,\ga^b]
\right\}\psi=0
\com
\label{ferKK5x}
\end{eqnarray}
%%%%%%%%%%%%%%%%%%%%%%%%%%%%%
See the upper half region of Fig.1. But this case reduces to (\ref{ferKK5})
by the transformation $\psi\ra\gago\psi$.
} 
We notice, in this expression, 
the EDM and  the MDM naturally appear\cite{Thirr72}.
They can be written as
%*** ferKK6 %%%%%%%%%%%%%%%%%%%%
\begin{eqnarray}
\frac{1}{16\m}eF_{ab}\psibar(c_E\ga^5+ic_M)[\ga^a,\ga^b]\psi\com\q
c_E\equiv\frac{\mhat}{M}\com\q c_M\equiv\frac{-\m}{M}
\pr
\label{ferKK6}
\end{eqnarray}
%%%%%%%%%%%%%%%%%%%%%%%%%%%%%
The first term(EDM) {\it violates} the CP symmetry.
The coefficient depends on the sign besides
the absolute value of $\mhat$. 
Writing the two coefficients as
$(c_E, c_M)\equiv (\mhat/M, -\m/M)$, we can consider
the following two limits:\ 
\begin{description}
\item[(i) CP-preserved limit [Small radius limit, 4D limit]]
\footnote{
If we regard the 4D charged fermion as the $N$-th KK mode,
$\m$ in (\ref{ferKK3}) is replaced by $N\m$. Then this CP-preserved
limit can also be regarded as {\it large N limit}. 
}
\nl
$\mhat/\m \ra \pm 0$, 
$(c_E, c_M)=(\pm 0, -1)$, $2\phi\ra 
-(\pi/2-0)$ [$3\pi/2-0$] for upper [lower] sign;\ 
\item[(ii) CP-extremely-violated limit [Large radius limit, 5D limit]]\nl
$\mhat/\m \ra \pm\infty$,  
$(c_E, c_M)=(\pm 1, 0)$, $2\phi\ra 
-0$ [$\pi+0$] for upper [lower] sign.
\end{description}  
Note that the MDM term usually appears, in the 4D QED, 
as the {\it quantum} effect.\nl
\nl
(B) Neutral Fermion\nl
We regard the neutral fermion as the {\it zero} mode of the KK-expansion.
%*** ferKK6b %%%%%%%%%%%%%%%%%%%%
\begin{eqnarray}
\psihat(x,y)=\psi(x)\com\nn
\left\{
\ga^a\pl_a
-\frac{f}{16}F_{ab}\ga^5[\ga^a,\ga^b]+\mhat
\right\}\psi(x)=0\pr
\label{ferKK6b}
\end{eqnarray}
%%%%%%%%%%%%%%%%%%%%%%%%%%%%%
Only the EDM term appears. 
This case is also realized by letting $e\ra 0,\ \m\ra 0$ 
keeping $e/\m=f$(fixed) in (\ref{ferKK5}).
Although the fermion has no charge, the
dipole moment appears. This case is similar to the limit (ii).\nl
\nl

Let us do the order estimation. From the reduction
%*** ferKK7 %%%%%%%%%%%%%%%%%%%%
\begin{eqnarray}
S=-\frac{1}{2G_5}\int d^5X\sqrt{-\gh}\Rhat=
-\frac{1}{2G_5}\frac{2\pi}{\m}\int d^4x 
\sqrt{-g}(R+\frac{f^2}{4}F^\ab F_\ab+\cdots)
\com
\label{ferKK7}
\end{eqnarray}
%%%%%%%%%%%%%%%%%%%%%%%%%%%%%
we know
%*** ferKK8 %%%%%%%%%%%%%%%%%%%%
\begin{eqnarray}
\frac{1}{G_5\m}\sim \frac{1}{G}\com\q
\frac{f^2}{G_5\m}\sim 1
\com
\label{ferKK8}
\end{eqnarray}
%%%%%%%%%%%%%%%%%%%%%%%%%%%%%
where $G$ is the (4D) gravitational constant. 
This gives $f\sim \sqrt{G}=10^{-19}\mbox{GeV}^{-1}$.
On the other hand, we know $e=\m f\sim 10^{-1}$.
Hence we obtain $\m\sim 10^{-1}f^{-1}\sim 10^{18}\mbox{GeV}$.
We originally have four parameters, $\mhat, \m, G_5$ and $f$. 
Among them we have the two relations (\ref{ferKK8}) from the
observation. We have another restriction among them from 
the present theoretical knowledge.
The most natural interpretation of the parameters is
that $\m^{-1}$ is the {\it infrared regularization}, $\mhat$ is
the energy scale of this 5D KK system. Then the validity
of the 5D {\it classical} treatment requires that
the 5D Planck mass $\gg |\mhat|$: 
%*** ferKK8b %%%%%%%%%%%%%%%%%%%%
\begin{eqnarray}
\frac{1}{\sqrt[3]{G_5}}\sim\sqrt[3]{100}\m
\gg |\mhat|\com
\label{ferKK8b}
\end{eqnarray}
%%%%%%%%%%%%%%%%%%%%%%%%%%%%%
and this reduces to, through the previous parameter
relations and values, $|\mhat|\ll 10^{19}$GeV. 

Now we consider the following three cases
in order to evaluate the EDM and MDM couplings.\newline
(0) $|\mhat|\sim\m$ \newline
We can estimate 
the electric and magnetic couplings as
%*** ferKK9 %%%%%%%%%%%%%%%%%%%%
\begin{eqnarray}
\frac{|\mhat|}{M\m}e
\sim\frac{e}{M}
%\sim 10^{-19}\mbox{GeV}^{-1}
\sim 10^{-32}~\mbox{e cm}
\pr
\label{ferKK9}
\end{eqnarray}
%%%%%%%%%%%%%%%%%%%%%%%%%%%%%
Both electric and magnetic moment terms
equally appear.
In this case, however, the theoretical restriction (\ref{ferKK8b})
is not so well satisfied.
\newline
(i) $|\mhat|\ll\m$(small radius, 4D limit),
%\footnote{
%small compared with the Compton wave length of the 5D fermion
%}
 CP-preserved limit\newline
We can estimate as
%*** ferKK10 %%%%%%%%%%%%%%%%%%%%
\begin{eqnarray}
\frac{\mhat}{M\m}e\sim\frac{\mhat}{\m}\times 10^{-32}~\mbox{e cm}\ ,\ 
\frac{e}{M}\sim\frac{e}{\m}\sim 10^{-32}~\mbox{e cm}\ ,\ 
|c_E|\ll |c_M|
\pr
\label{ferKK10}
\end{eqnarray}
%%%%%%%%%%%%%%%%%%%%%%%%%%%%%
In this case the EDM coupling is suppressed by the 
factor of the mass parameter ratio $\frac{|\mhat|}{\m}$($\ll 1$).
The theoretical restriction (\ref{ferKK8b})
is satisfied, hence this parameter region is well controlled
theoretically.
\newline
(ii) $|\mhat|\gg\m$(large radius, 5D limit), 
CP-extremely-violated limit \newline
We can estimate as
%*** ferKK11 %%%%%%%%%%%%%%%%%%%%
\begin{eqnarray}
\frac{|\mhat|}{M\m}e\sim\frac{e}{\m}\sim 10^{-32}~\mbox{e cm}\com\q
\frac{e}{M}\sim\frac{\m}{|\mhat|}\times 10^{-32}~\mbox{e cm}\ ,\ 
|c_E|\gg |c_M|
\pr
\label{ferKK11}
\end{eqnarray}
%%%%%%%%%%%%%%%%%%%%%%%%%%%%%
In this case the MDM coupling is suppressed by the  
factor $\frac{\m}{|\mhat|}$($\ll 1$).
The theoretical restriction (\ref{ferKK8b}), however,
is not satisfied. This implies the 5D {\it quantum} effect can not be
negligible in this parameter region.

We note that the ratio of 
the two massive parameters, 
(the radius of the extra space)$^{-1}$ $\m$ and 
the 5D fermion mass $\mhat$, controls
the {\it dual} aspect (electric versus magnetic) of the theory.
This point will be compared with the RS case later
(see the ending paragraphs of Sec.7). 

As for the EDM,
all cases are {\it far below} the experimental upper bound, for example, 
the neutron EDM is less than $6.3\times 10^{-26}$e cm\cite{Harris99}.
As for the MDM,  
we know, (from the formula of the quantum effect of 4D QED: 
$e\hbar/2mc$,) the order
of the observed values are
$10^{-11}$ e~cm for the electron, 
$10^{-14}$ e~cm for the proton,
$10^{-16}$ e~cm for the top quark.
The prediction of 5D KK theory is {\it superweak} compared with these values. 
Hence the present model is viable %with the experimental data 
but quantitatively not so attractive.
\footnote{
The 5D KK model has another phenomenological defect in the prediction of
the 4D charged fermion mass. The order of the magnitude is
$M=\sqrt{\mhat^2+\m^2}\sim\m\sim 10^{18}$ GeV, which
cannot be accepted as the ordinary (charged) fermion masses.
From this, the MDM value essentially does not improve
even if the quantum effect (of the 4D QED with the Planck mass fermion)
is taken into account. 
}
 We are, at present, 
content with the qualitatively interesting point. 

Thirring\cite{Thirr72} showed that 
the CP-violating term appears
not because the discrete symmetries
(charge conjugation, parity and time reversal) 
do not exist in 
the Dirac equation (\ref{ferKK5}), 
%which is reduced from the 5D theory (\ref{ferKK1}),  
but because they
appear in the form which differs from the ordinary one.

%%%%%%%%%%%%%%%%%%%%%%%%%%%%%%%%%%%%%%%%%%%%%%%%%%%%%%%%%%%%%%%%%%%%%
%%%% SEC. 4  Randall-Sundrum Theory in the Cartan Formalism %%%%%%%%%
%%%%%%%%%%%%%%%%%%%%%%%%%%%%%%%%%%%%%%%%%%%%%%%%%%%%%%%%%%%%%%%%%%%%%
\section{Randall-Sundrum Theory in the Cartan Formalism}
Let us formulate the RS theory in the Cartan formalism.
We consider the following 5D space-time geometry\cite{RS9905,RS9906}.
%In terms of the 4D coordinates $x^a$ and the extra one
%$y$,  the metric is taken as follows. 
%(We also use the notation $(X^m)=(x^a,y)$. ) 
%*** RS1 %%%%%%%%%%%%%%%%%%%%
\begin{eqnarray}
ds^2=\e^{-2\si(y)}\eta_{ab}dx^adx^b+dy^2
=\gh_{mn}dX^mdX^n\com\nn
-\infty<y<+\infty\ ,\ -\infty<x^a<+\infty
\com
\label{RS1}
\end{eqnarray}
%%%%%%%%%%%%%%%%%%%%%%%%%%%%%
where $\si(y)$ is a "scale factor" field.
$(\eta_{ab})=\mbox{diag}(-1,1,1,1)$.
When the geometry is AdS$_5$, $\si(y)=c|y|, c>0$. 
[ Such a situation, in the present case, occurs in the asymptotic region
of the extra space $y\ra\pm\infty$.\cite{SI0003,SI0107}] 
Note that the "scale factor" field $\si(y)$ depends
only on the extra coordinate $y$ and controls the 4D metric
part $\eta_{ab}dx^adx^b$, whereas the corresponding field
$\si(x)$ in the KK case of (\ref{KK1c}) depends only on
the 4D coordinates
$x^a$ and mainly controls the extra part $(dy)^2$. 
%Note that the 4D world $\{x^a\}$ and the extra space
%$\{y\}$ is {\it not} decomposable. (cf KK case with $\si=0$.)
As in Sec.2, we introduce the local Lorentz frame as
%*** RS2 %%%%%%%%%%%%%%%%%%%%
\begin{eqnarray}
ds^2=\thhat^\m\thhat^\n \etahat_\mn
\com\q 
(\etahat_\mn)=\mbox{diag}(-1,1,1,1,1)
\pr
\label{RS2}
\end{eqnarray}
%%%%%%%%%%%%%%%%%%%%%%%%%%%%%
The 1-form $\thhat^\m$, the basis
of the cotangent manifold $T_p^*M$, is given by 
%*** RS3 %%%%%%%%%%%%%%%%%%%%
\begin{eqnarray}
\thhat^\al=\e^{-\si}\eta^\al_{~a}dx^a\com\q
\thhat^5=dy
\com
\label{RS3}
\end{eqnarray}
%%%%%%%%%%%%%%%%%%%%%%%%%%%%%
from which we can read off the f\"{u}nf-bein 
$\ehat^\m_{~m}$ and its inverse $\ehat_\m^{~m}$ as
%*** RS4 %%%%%%%%%%%%%%%%%%%%
\begin{eqnarray}
(\ehat^\m_{~m})=\left(
\begin{array}{cc}
e^{-\si}\eta^\al_{~a} & 0 \\
0   & 1
\end{array}
                 \right)\com\q
(\ehat_\m^{~m})=\left(
\begin{array}{cc}
e^\si\eta_\al^{~a} & 0 \\
0   & 1
\end{array}      \right)
                 \pr
\label{RS4}
\end{eqnarray}
%%%%%%%%%%%%%%%%%%%%%%%%%%%%%
Taking 1-form $\thhat^\m$ of (\ref{RS3}), and
compare the results with the 1st Cartan's structure
equation (\ref{KK8}), we obtain the connection 1-form
$\omhat^\m_{~\n}$ as
%*** RS5 %%%%%%%%%%%%%%%%%%%%
\begin{eqnarray}
\omhat^\fbar_{~\fbar}=0\com\q 
\omhat^\al_{~\fbar}=-\omhat_\fbar^{~\al}=-\si'\th^\al\com\q
\omhat^\fbar_{~\al}=-\omhat_\al^{~\fbar}=\si'\th_\al\com\q
\omhat^\al_{~\be}=0
\com
\label{RS5}
\end{eqnarray}
%%%%%%%%%%%%%%%%%%%%%%%%%%%%%
where $\si'=\frac{d\si}{dy}$.

Further explanation of the RS geometry in the Cartan formalism
is done in App.B where
a more general coordinate is considered.
%2nd Cartan, Riemann curvature ***********
%%%%%%%%%%%%%%%%%%%%%%%%%%%%%%%%%%%%%%%%%%%%%%%%%%%%%%%%%%%%%%%%%%%%
%%%% SEC. 5  Fermions in Randall-Sundrum Theory  %%%%%%%%%%%%%%%
%%%%%%%%%%%%%%%%%%%%%%%%%%%%%%%%%%%%%%%%%%%%%%%%%%%%%%%%%%%%%%%%%%%%%
\section{Fermions in Randall-Sundrum Theory}
The 5D Dirac Lagrangian in the RS theory is given, 
from (\ref{ferKK1}) and the results of Sec.4, as
%*** ferRS1 %%%%%%%%%%%%%%%%%%%%
\begin{eqnarray}
\sqrt{-\gh}\Lcal^{Dirac}=\sqrt{-\gh}i{\bar \psihat}
\left\{
\ga^\m\ehat_\m^{~m}\pl_m
+\frac{1}{8}(\omhat^\si)_\mn\ga_\si [\ga^\m,\ga^\n]+\mhat(y)
\right\}\psihat                             \nn
=i\e^{-3\si}{\bar \psihat}\{
\ga^a\pl_a-2\e^{-\si}(\si'-\half\pl_y)\gago
               +\mhat(y) e^{-\si}     \}\psihat\nn
=i\e^{-\frac{3}{2}\si}{\bar \psihat}\{
\ga^a\pl_a-2\e^{-\si}(\fourth\si'-\half\pl_y)\gago
                      +\mhat(y) e^{-\si}  \}(\e^{-\frac{3}{2}\si}\psihat)
\com
\label{ferRS1}
\end{eqnarray}
%%%%%%%%%%%%%%%%%%%%%%%%%%%%%
where $\mhat$ is the 5D fermion mass $-\infty<\mhat<+\infty$.
For the later use, we here allow $\mhat$ to have the $y$-dependence :\ 
$\mhat=\mhat(y)$. 
\footnote{
The situation is realized by the bulk Higgs mechanism treated in Sec.6.
}
The special case of this result, $\mhat=0$, coincides with an equation cited in \cite{CHNOY99}. 
%\footnote{
%The 5D fermion mass term, $im{\bar \psihat}\psihat$, 
%is a pseudo scalar(5D parity is odd). This will be 
%corrected, in the next section, by introducing
%the 5D pseudo scalar (Higgs) $\Phi$ and by replacing
%the mass term by $i\Phi{\bar \psihat}\psihat$. 
%}

Let us do the dimensional reduction from 5D to 4D
\cite{KS00,GN99,CHNOY99}.
We take the following form of expansion.
%*** ferRS2 %%%%%%%%%%%%%%%%%%%%
\begin{eqnarray}
\psihat(x,y)=\sum_k (\psi^k_L(x)\xi_k(y)+\psi^k_R(x)\eta_k(y)),\nn
\gago\psi_L(x)=-\psi_L(x)\com\q 
\gago\psi_R(x)=+\psi_R(x)
\com
\label{ferRS2}
\end{eqnarray}
%%%%%%%%%%%%%%%%%%%%%%%%%%%%%
where $\{\xi_k(y),\eta_k(y)\}$ is a complete set of some
eigenfunctions to be determined. This expansion
corresponds to (\ref{KK2}) and (\ref{ferKK3}) in
the Kaluza-Klein case. 
The role of the periodic eigenfunctions $\{\e^{ik\m y}\}$
is here played by $\{\xi_k(y),\eta_k(y)\}$.
For simplicity, we consider
the "5D-{\it parity}" even case for $\psihat(x,y)$.
%*** ferRS3 %%%%%%%%%%%%%%%%%%%%
\begin{eqnarray}
\gago\psihat(x,-y)=+\psihat(x,y)
\pr
\label{ferRS3}
\end{eqnarray}
%%%%%%%%%%%%%%%%%%%%%%%%%%%%%
(The odd case is similarly examined.)
This requires $\xi_n(y)$ to be an {\it odd} function and
$\eta_n(y)$ to be an {\it even} function with respect to
the {\it $Z_2$ transformation}:\ $y\change -y$. 
%*** ferRS4 %%%%%%%%%%%%%%%%%%%%
\begin{eqnarray}
\xi_n(-y)=-\xi_n(y)\com\q \eta_n(-y)=+\eta_n(y)
\pr
\label{ferRS4}
\end{eqnarray}
%%%%%%%%%%%%%%%%%%%%%%%%%%%%%
From these we get the following important boundary
conditions,
%*** ferRS4b %%%%%%%%%%%%%%%%%%%%
\begin{eqnarray}
\xi_n(0)=0\q\mbox{(Dirichlet)}\com\q 
\pl_y\eta_n|_{y=0}=0\q\mbox{(Neumann)}
\com
\label{ferRS4b}
\end{eqnarray}
%%%%%%%%%%%%%%%%%%%%%%%%%%%%%
when $\xi_n(y)$ and $\pl_y\eta_n(y)$ are continuous
at $y=0$. 
The Lagrangian reduces to
%*** ferRS5 %%%%%%%%%%%%%%%%%%%%
\begin{eqnarray}
\sqrt{-\gh}\Lcal^{Dirac}
=i\sum_m(\psibar^m_L\xitil_m(y)+\psibar^m_R\etatil_m(y))\times\nn
\sum_n\{
\ga^a\pl_a\psi^n_L\xitil_n-
\e^{-\si}(\frac{\si'}{2}-\pl_y)(-\psi^n_L)\xitil_n+
\ga^a\pl_a\psi^n_R\etatil_n-
\e^{-\si}(\frac{\si'}{2}-\pl_y)\psi^n_R\etatil_n      \nn
+\e^{-\si}\mhat(y) (\psi^n_L\xitil_n+\psi^n_R\etatil_n)
  \}
\com
\label{ferRS5}
\end{eqnarray}
%%%%%%%%%%%%%%%%%%%%%%%%%%%%%
where we define $\xitil_n\equiv \e^{-\frac{3}{2}\si}\xi_n$ and
$\etatil_n\equiv \e^{-\frac{3}{2}\si}\eta_n$.

We now take the set of eigenfunctions $\{\xitil_n,\etatil_n\}$ as
%*** ferRS6 %%%%%%%%%%%%%%%%%%%%
\begin{eqnarray}
\e^{-\si}(\frac{\si'}{2}-\pl_y)\xitil_n
+\e^{-\si}\mhat(y)\xitil_n=m_n\etatil_n\ ,\ 
-\e^{-\si}(\frac{\si'}{2}-\pl_y)\etatil_n
+\e^{-\si}\mhat(y)\etatil_n=m_n\xitil_n
\ ,
\label{ferRS6}
\end{eqnarray}
%%%%%%%%%%%%%%%%%%%%%%%%%%%%%
which are orthnormalized as
%*** ferRS7 %%%%%%%%%%%%%%%%%%%%
\begin{eqnarray}
\int^\infty_{-\infty}dy~\xitil_n(y)\xitil_m(y)=
\int^\infty_{-\infty}dy~\etatil_n(y)\etatil_m(y)=\del_{nm}\com\nn
\int^\infty_{-\infty}dy~\xitil_n(y)\etatil_m(y)=0
\pr
\label{ferRS7}
\end{eqnarray}
%%%%%%%%%%%%%%%%%%%%%%%%%%%%%
Then the 5D action (\ref{ferRS1}) finally reduces to the sum of 4D {\it free}
fermions.
%*** ferRS8 %%%%%%%%%%%%%%%%%%%%
\begin{eqnarray}
\int d^5X\sqrt{-\gh}\Lcal^{Dirac}=
i\int d^4x\sum_n\{
\psibar^n_L(\ga^a\pl_a\psi^n_L+m_n\psi^n_R)+
\psibar^n_R(\ga^a\pl_a\psi^n_R+m_n\psi^n_L)
                \}
\ .
\label{ferRS8}
\end{eqnarray}
%%%%%%%%%%%%%%%%%%%%%%%%%%%%%
The information of this fermion dynamics is now in 
the set of the eigen values
$\{m_n\}$ determined by (\ref{ferRS6}). 

From the coupled equation (\ref{ferRS6}) with respect to
$\xitil_n$ and $\etatil_n$, we get
the differential equation for $\xitil_n$ as
%*** ferRS9 %%%%%%%%%%%%%%%%%%%%
\begin{eqnarray}
\e^{-2\si}[\frac{\si''}{2}-\frac{3}{4}{\si'}^2
+2\si'\pl_y-\mhat(y)\si'+{\mhat(y)}^2
-{\pl_y}^2+{\mhat(y)}']\xitil_n={m_n}^2\xitil_n
\pr
\label{ferRS9}
\end{eqnarray}
%%%%%%%%%%%%%%%%%%%%%%%%%%%%%
(We can obtain a similar one for $\etatil_n$.)
For simplicity, we consider the {\it thin wall limit}:
%*** ferRS10 %%%%%%%%%%%%%%%%%%%%
\begin{eqnarray}
\si(y)=\om|y| \com\q \si'(y)=\om\ep(y)\com\q
\si''(y)=2\om\del(y)\com\nn
\q\q\q\q\q\mhat(y)=\mtil\ep(y)\com\q\mhat'(y)=2\mtil\del(y)
\com
\label{ferRS10}
\end{eqnarray}
%%%%%%%%%%%%%%%%%%%%%%%%%%%%%
where $\om(>0)$ and $\mtil(>0)$ are some constants.
$\ep(y)$ is the sign function:\ 
$\ep(y)=1$ for $y>0$ and $\ep(y)=-1$ for $y<0$
\ ($\ep'(y)=2\del(y)$). 
In this
limit, the equation (\ref{ferRS9}) can be explicitly
solved. 
%*** ferRS11 %%%%%%%%%%%%%%%%%%%%
\begin{eqnarray}
\e^{-2\om|y|}[(\om+2\mtil)\del(y)-\frac{3}{4}\om^2
+2\om\ep(y)\pl_y-\mtil\om+\mtil^2
-{\pl_y}^2]\xitil_n={m_n}^2\xitil_n
\com
\label{ferRS11}
\end{eqnarray}
%%%%%%%%%%%%%%%%%%%%%%%%%%%%%
where ${\ep(y)}^2=1$ is used. The presence of $\del(y)$ indicates 
a singularity of the solution at $y=0$. 
Let us see the solution in the region $y>0$. 
In terms of a new coordinate $z\equiv\frac{1}{\om}\e^{\om y}$,
the above equation reduces to the {\it Bessel differential
equation}.  
%*** ferRS12 %%%%%%%%%%%%%%%%%%%%
\begin{eqnarray}
\{{\pl_z}^2-\frac{1}{z}\pl_z+\frac{1-\n^2}{z^2}
+{m_n}^2\}\xitil_n=0\com\nn
\n=|\frac{\mtil}{\om}-\half|
\pr
\label{ferRS12}
\end{eqnarray}
%%%%%%%%%%%%%%%%%%%%%%%%%%%%%
The solution $\xitil_n$ is obtained as
%*** ferRS13 %%%%%%%%%%%%%%%%%%%%
\begin{eqnarray}
\xitil_n(y)=\frac{1}{(\om z)^{3/2}}\xi_n(z)
=z\{ J_\n(m_nz)+c_nN_\n(m_nz)\}\com\nn
c_n=-\frac{J_\n(m_n/\om)}{N_\n(m_n/\om)}
\com
\label{ferRS13}
\end{eqnarray}
%%%%%%%%%%%%%%%%%%%%%%%%%%%%%
where $c_n$ is determined by the Dirichlet boundary
condition (\ref{ferRS4b}). $J_\n(z)$ and $N_\n(z)$
are two independent Bessel functions. With the above
explicit solution of $\xitil_n$, we can obtain $\etatil_n$
using the first equation of (\ref{ferRS6}).
Another boundary condition (Neumann) on $\eta_n$ (\ref{ferRS4b})
gives us the set $\{m_n\}$ as the zeros of some combination
of Bessel functions.
(
For the special cases $\mtil=0$ (no 5D fermion mass)
or $\mtil/\om=1$, the eigen functions reduce to
$J_{1/2}(z)=\sqrt{2/\pi z}\sin~z,\ 
N_{1/2}(z)=-J_{-1/2}(z)=-\sqrt{2/\pi z}\cos~z$.
The former case is that one considered in \cite{CHNOY99}.
In these cases, however, the fermion localization does not
occur in the present one wall model. See Sec.6.  
)

We now understand that the periodic eigen functions
$\{\e^{ik\m y}\}$ in the KK case ((\ref{KK2}) and (\ref{ferKK3}))
correspond to $\{\xi_k(y),\eta_k(y)\}$ specified by (\ref{ferRS6}).
Only for the thin wall limit, they are explicitly solved
by Bessel functions. Although we do not explicitly require the compactness
of the extra space, the eigen values are $S^1$-like. This is
due to the requirement of the geometry AdS$_5$($\approx$S$^1\times$R$^4$) 
in the asymptotic region. 
%the $Z_2$ symmetry requirement (\ref{ferRS3}). 

The importance of the 5D mass "function" $\mhat(y)$ is now clear.
In the next section, we explain  its origin in the bulk field theory.
Nature requires the {\it Yukawa interaction} 
between the 5D fermion and the 5D Higgs\cite{BG99}.

%%%%%%%%%%%%%%%%%%%%%%%%%%%%%%%%%%%%%%%%%%%%%%%%%%%%%%%%%%%%%%%%%%%%%
%%%%      SEC. 6  Bulk Higgs Mechanism             %%%%%%%%%%%%%%%%%%
%%%%              in the Randall-Sundrum Geometry  %%%%%%%%%%%%%%%%%%
%%%%%%%%%%%%%%%%%%%%%%%%%%%%%%%%%%%%%%%%%%%%%%%%%%%%%%%%%%%%%%%%%%%%%
\section{Bulk Higgs Mechanism 
%in the Randall-Sundrum Geometry
}
One of the most important characters of the brane world model
is the massless chiral fermion localization. It is phenomenologically
attractive because the smallness of the quark and lepton masses
could be naturally explained. Theoretically it is also necessary
as the dimensional reduction mechanism. The feature comes from
the $Z_2$ ($y\change -y$) properties of the system. 
The most natural way to introduce the properties
is to use the
{\it Higgs mechanism} in the bulk world. 
\footnote
{In the case of the flat space-time, 
Rubakov and Shaposhnikov\cite{RS83}
proposed a domain wall model
caused by the bulk Higgs potential.
}

Let us examine the case the fermion system has
the Yukawa coupling with the bulk Higgs field.
%*** yuka1 %%%%%%%%%%%%%%%%%%%%
\begin{eqnarray}
\sqrt{-\gh}\Lcal=\sqrt{-\gh}(\Lcal^{Dirac}+\Lcal^Y)\com\q
\Lcal^Y=ig_Y{\bar \psihat}\psihat\Phi
\com
\label{yuka1}
\end{eqnarray}
%%%%%%%%%%%%%%%%%%%%%%%%%%%%%
where the Higgs field $\Phi$ is the 5D(bulk) scalar field
and $g_Y$ is the Yukawa coupling.
%**  goto the biggining of Sec.4 ???  ** 
We assume that the Higgs field, besides 
the "scale factor" field $\si(y)$, 
is some background given by 
the (classical) field equation of
the 5D gravity-Higgs system.
%*** yuka2 %%%%%%%%%%%%%%%%%%%%
\begin{eqnarray}
\sqrt{-\gh}(\Lcal^{grav}+\Lcal^S)\com\nn
\Lcal^{grav}=\frac{-1}{2G_5}{\hat R}\com\q
\Lcal^{S}=-\half\na_m\Phi\na^m\Phi-V(\Phi)\com\q
\com
\label{yuka2}
\end{eqnarray}
%%%%%%%%%%%%%%%%%%%%%%%%%%%%%
where 
$V(\Phi)$ is the ordinary Higgs potential.
In ref.\cite{SI0003,SI0107}, it is shown that the above
gravity-Higgs system has a {\it stable} kink (domain wall)
solution for the case $\Phi=\Phi(y)$. 
In the IR asymptotic region from the wall,
$\si'(y)$ and $\Phi(y)$ behave as
%*** yuka3 %%%%%%%%%%%%%%%%%%%%
\begin{eqnarray}
\si'(y)=\left\{\begin{array}{c}+\om,\ \ ky\ra +\infty\\
                               -\om,\ \ ky\ra -\infty
                \end{array}\right.\com\q
\Phi(y)=\left\{
\begin{array}{c}
+\vz,\ \ ky\ra +\infty \\
-\vz,\ \ ky\ra -\infty

\end{array}
                 \right.                 
\com\label{yuka3}
\end{eqnarray}
%%%%%%%%%%%%%%%%%%%%%%%%%%%%%
where $k$(the inverse of thickness), $\om$(brane tension)
%(brane tension ?***) 
and $\vz$(5D Higgs vacuum expectation value) are some positive
constants expressed by a free parameter, the vacuum parameters
and the 5D gravitational constant. Near the origin of the extra axis
($k|y|\ll 1$), they behave as
%*** yuka4 %%%%%%%%%%%%%%%%%%%%
\begin{eqnarray}
\si'(y)=\om\tanh (ky)
\com\q
\Phi(y)=\vz\tanh (ky)
\pr
\label{yuka4}
\end{eqnarray}
%%%%%%%%%%%%%%%%%%%%%%%%%%%%%

The dimensional reduction to 4D is performed by
taking the the {\it thin wall} limit $k\ra \infty$, 
which is precisely defined as
%*** yuka4b %%%%%%%%%%%%%%%%%%%%%
\begin{eqnarray}
k\gg \frac{1}{r_c}
\pr
\label{yuka4b}
%\label{qed8}
\end{eqnarray}
%%%%%%%%%%%%%%%%%%%%%%%%%%%%%
where $r_c$ is the {\it infrared cutoff} 
of the extra axis ($-r_c<y<r_c$). (See ref.\cite{SI0003}.)
In this limit, above quantities behave as 
$\si'(y)\ra \om\th(y),\ \Phi(y)\ra \vz\th(y)$.  
%**  goto the biggining of Sec.4 ???  ** 
All dimensional parameters are
a) ${G_5}^{-1/3}$:\ 5D Planck mass;\ 
b) $|\mhat|=g_Y\vz$:\ 5D fermion mass;\ 
c) $k^{-1}$:\ thickness of the domain;\ 
d) $r_c$:\ Infrared regularization of the extra axis.
Among them  there exists a theoretical restriction
from the requirement:\ 
5D {\it classical} treatment works well.
%*** yuka4c %%%%%%%%%%%%%%%%%%%%
\begin{eqnarray}
\frac{1}{\sqrt[3]{G_5}}\gg k\com
\label{yuka4c}
\end{eqnarray}
%%%%%%%%%%%%%%%%%%%%%%%%%%%%%

The 5D Dirac equation of (\ref{yuka1})
 is given by (cf. eq.(\ref{ferRS1})),
%*** yuka5 %%%%%%%%%%%%%%%%%%%%
\begin{eqnarray}
i\e^{\si}\{
\ga^a\pl_a-2\e^{-\si}(\si'-\half\pl_y)\gago+g_Y\e^{-\si}\Phi
                           \}\psihat=0
\pr
\label{yuka5}
\end{eqnarray}
%%%%%%%%%%%%%%%%%%%%%%%%%%%%%
This is just the lagragian of (\ref{ferRS1})
with $\mhat(y)=g_Y\Phi(y)$. 
Let us examine a solution of the right chirality zero mode.
%*** yuka6 %%%%%%%%%%%%%%%%%%%%
\begin{eqnarray}
\psihat(x,y)=\psi^0_R(x)\eta(y)\com\q
\gago\psi^0_R=+\psi^0_R\com\q
\ga^a\pl_a\psi^0_R=0
\pr
\label{yuka6}
\end{eqnarray}
%%%%%%%%%%%%%%%%%%%%%%%%%%%%%
The equation (\ref{yuka5}) reduces to
%*** yuka7 %%%%%%%%%%%%%%%%%%%%
\begin{eqnarray}
\pl_y\eta=(2\si'+g_Y\Phi(y))\eta
\pr
\label{yuka7}
\end{eqnarray}
%%%%%%%%%%%%%%%%%%%%%%%%%%%%%
In the IR asymptotic region($k|y|\gg 1$), 
the solution behaves as
%*** yuka8 %%%%%%%%%%%%%%%%%%%%
\begin{eqnarray}
\eta(y)=\mbox{const}\times \e^{-(g_Y\vz-2\om)|y|}
\com
\label{yuka8}
\end{eqnarray}
%%%%%%%%%%%%%%%%%%%%%%%%%%%%%
which shows the {\it exponentially damping} for the large
yukawa coupling $g_Y>2\om/\vz$\cite{BG99}. This is called
{\it massless chiral fermion localization}. Near the origin
of the extra axis($k|y|\ll 1$), $\eta(y)$ behaves as
%*** yuka9 %%%%%%%%%%%%%%%%%%%%
\begin{eqnarray}
\eta(y)=\mbox{const}\times \e^{-\frac{k}{2}(g_Y\vz-2\om)y^2}
\com
\label{yuka9}
\end{eqnarray}
%%%%%%%%%%%%%%%%%%%%%%%%%%%%%
which shows the {\it Gaussian damping}. The behavior is shown
in Fig.2. It shows the {\it regular} property of the solution.
For the left chirality zero-mode, we can show the same behavior
using the anti-kink solution instead of the kink solution (\ref{yuka3}). 
%**  General modes, Bessel equation   ****
%%%%%%%%%%%%%%%%%  Fig.2 %%%%%%%%%%%%%%%%%%%%%
\begin{figure}%[htb]
\epsfysize=3cm\epsfbox{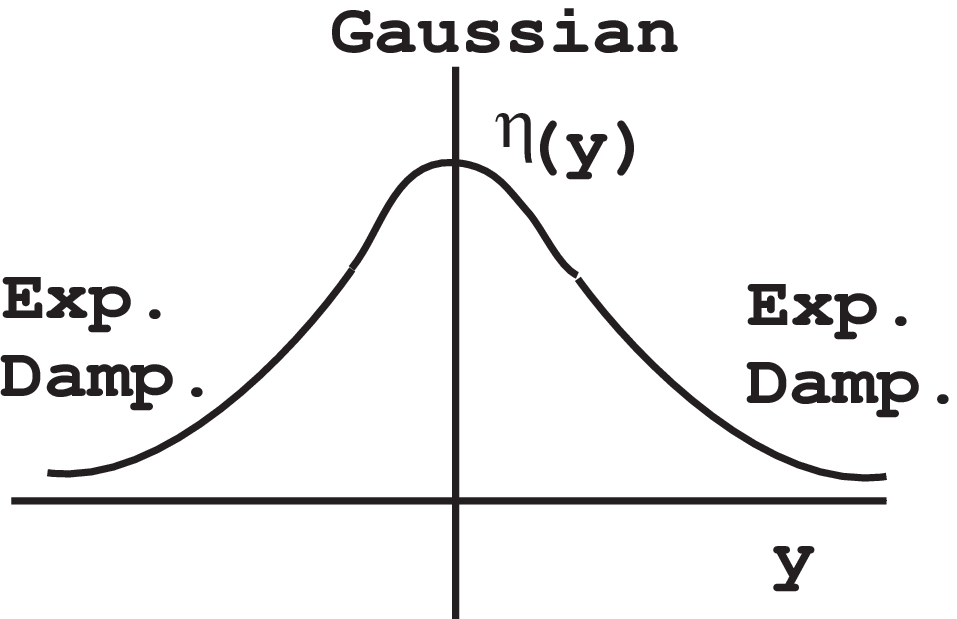}
   \begin{center}
Fig.2\ Behavior of the fermion along the extra axis.
%***local2.eps
   \end{center}
\label{fig:local2}
\end{figure}
%%%%%%%%%%%%%%%%%%%%%%%%%%%%%%%%%%%%%%%%%%%%%%%

%%%%%%%%%%%%%%%%%%%%%%%%%%%%%%%%%%%%%%%%%%%%%%%%%%%%%%%%%%%%%%%%%%%%%
%%%%      SEC. 7  5D QED                           %%%%%%%%%%%%%%%%%%
%%%%              in the Randall-Sundrum Geometry  %%%%%%%%%%%%%%%%%%
%%%%%%%%%%%%%%%%%%%%%%%%%%%%%%%%%%%%%%%%%%%%%%%%%%%%%%%%%%%%%%%%%%%%%
\section{Five Dimensional QED 
%in the Randall-Sundrum Geometry 
and Bulk Quantum Effect}
Let us examine the 5D QED, 
$\Lcal^{QED}=-e{\bar \psihat}
\ga^\m\ehat_\m^{~m}\psihat A_m$, 
with the Yukawa interaction in RS geometry.
\footnote{
The U(1) gauge space is here an internal space:\ 
$A_m\ra A_m+\pl_m\La(X), \psihat\ra \exp(-ie\La(X))~\psihat$.
This should be compared with the case of KK where the U(1)
gauge space is identified with the extra (y-) space. 
}
%*** qed1 %%%%%%%%%%%%%%%%%%%%
\begin{eqnarray}
\sqrt{-\gh}(\Lcal^{Dirac}+\Lcal^{QED}+\Lcal^Y)\nn
=\sqrt{-\gh}\left[ i{\bar \psihat}
\left\{
\ga^\m\ehat_\m^{~m}(\pl_m+ieA_m)
+\frac{1}{8}(\omhat^\si)_\mn\ga_\si [\ga^\m,\ga^\n]
\right\}\psihat+ig_Y{\bar \psihat}\psihat\Phi
                \right]
\pr
\label{qed1}
\end{eqnarray}
%%%%%%%%%%%%%%%%%%%%%%%%%%%%%
The kinetic (propagator) part for
the electromagnetic, gravitational and Higgs fields is 
provided by 
%*** qed2 %%%%%%%%%%%%%%%%%%%%
\begin{eqnarray}
\sqrt{-\gh}(\Lcal^{EM}+\Lcal^{grav}+\Lcal^S)\com\q
\Lcal^{EM}=-\fourth \gh^{mn}\gh^{kl}F_{mk}F_{nl}
\com
\label{qed2}
\end{eqnarray}
%%%%%%%%%%%%%%%%%%%%%%%%%%%%%
where $\Lcal^{grav},\Lcal^S$ are given in (\ref{yuka2}). 
We assume, as in Sect.5 and 6, $\gh^{mn}$ and $\Phi$ are
the brane background fields obtained as the stable
solution of the system $\Lcal^{grav}+\Lcal^S$.

We have introduced the Yukawa coupling in order to localize
the fermion on the wall. This model, however, is still 
unsatisfactory in that 
the vector (gauge) field is {\it not localized}\cite{BG99}. One resolution
is to take 6D model\cite{Oda00}. Here we are content only with
the fermion part and do not pursue this 
%vector **delocalization**
problem.

Let us examine the {\it bulk quantum} effect. It induces the 5D 
effective action $S_{eff}$ which reduces to a 4D action
in the {\it thin wall} limit. From the diagram of Fig.3, we expect
%*** qed3 %%%%%%%%%%%%%%%%%%%%
\begin{eqnarray}
\frac{\del S^{(1)}_{eff}}{\del A^\mu(X)}\equiv
<J_\m>\sim e^2g_Y\ep_{\mn\ls\tau}\Phi F^{\nu\la}F^{\si\tau}
\pr
\label{qed3}
\end{eqnarray}
%%%%%%%%%%%%%%%%%%%%%%%%%%%%%
%%%%%%%%%%%%%%%%%  Fig.3 %%%%%%%%%%%%%%%%%%%%%
\begin{figure}%[htb]
\epsfysize=3cm\epsfbox{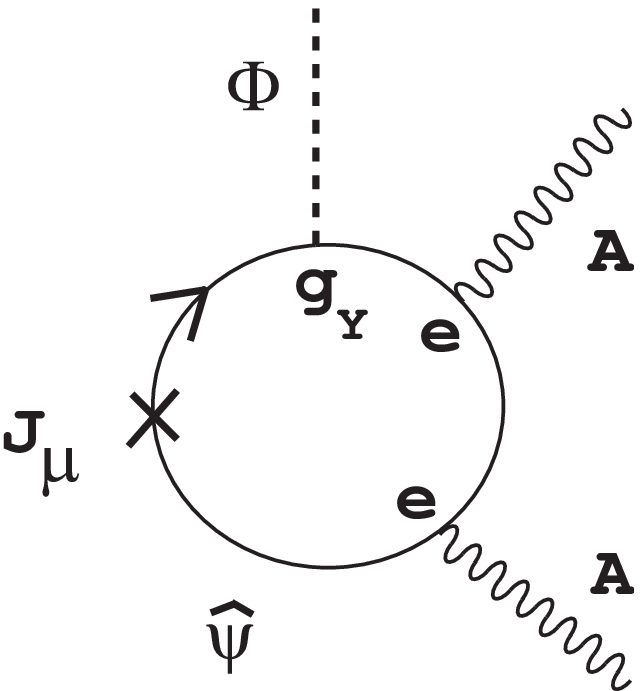}
   \begin{center}
Fig.3\ A bulk quantum-loop diagram. The diagram induces the effective action
$S^{(1)}_{eff}$.
%***anomaly2.eps
   \end{center}
\label{fig:anomaly2}
\end{figure}
%%%%%%%%%%%%%%%%%%%%%%%%%%%%%%%%%%%%%%%%%%%%%%%
Then the effective action is integrated as
%*** qed4 %%%%%%%%%%%%%%%%%%%%
\begin{eqnarray}
S^{(1)}_{eff}\sim
e^2g_Y\int d^5X\ep_{\mn\ls\tau}\Phi A^\m F^{\nu\la}F^{\si\tau}
\pr
\label{qed4}
\end{eqnarray}
%%%%%%%%%%%%%%%%%%%%%%%%%%%%%
%(****When $\Phi$=const., this is the 5D Chern-Simon term.****)
In the {\it thin wall limit} we may approximate as
$\Phi=\Phi(y)\sim \vz\ep(y)$ where $\ep(y)$ is the step function.
(See the description below (\ref{yuka4}).)
Under the U(1) gauge transformation $\del A^\m=\pl^\m\La$, 
$S^{(1)}_{eff}$ changes as
%*** qed5 %%%%%%%%%%%%%%%%%%%%
\begin{eqnarray}
\del_\La S^{(1)}_{eff}\sim
e^2g_Y\vz\int d^5X\ep_{\mn\ls\tau}\ep(y) \pl^\m\La
F^{\nu\la}F^{\si\tau}\nn
=e^2g_Y\vz\int d^5X\{
\pl^\m(\ep_{\mn\ls\tau}\ep(y) \La F^{\nu\la}F^{\si\tau})
-\ep_{5\nu\ls\tau}\del(y) \La F^{\nu\la}F^{\si\tau}
                    \}\nn
=-e^2g_Y\vz\int d^4x
\La(x) F^{\ab}{\tilde F}_{\ab}
\com
\label{qed5}
\end{eqnarray}
%%%%%%%%%%%%%%%%%%%%%%%%%%%%%
where ${\tilde F}_{\ab}\equiv \ep_{\ab\ga\del}F^{\ga\del}$. 
In the above we assume that the boundary term vanishes. 
Callan and Harvey interpreted this result as the "anomaly flow"
between the boundary (our 4D world) and the bulk\cite{CH85}. 
Through the analysis of 
the {\it induced action} in the bulk,
we can see the {\it dual} aspect of the 4D QED. 

Another interesting bulk quantum effect is given by Fig.4.
%%%%%%%%%%%%%%%%%  Fig.4 %%%%%%%%%%%%%%%%%%%%%
\begin{figure}%[htb]
\epsfysize=3cm\epsfbox{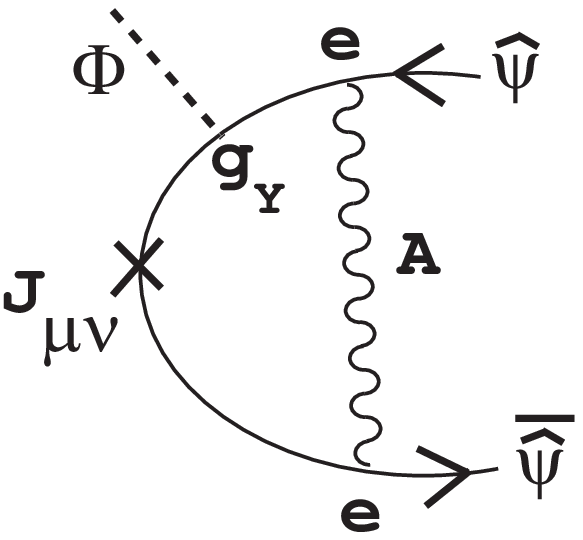}
   \begin{center}
Fig.4\ A bulk quantum-loop diagram. The diagram induces the effective action
$S^{(2)}_{eff}$.
%***magne2.eps
   \end{center}
\label{fig:magne2}
\end{figure}
%%%%%%%%%%%%%%%%%%%%%%%%%%%%%%%%%%%%%%%%%%%%%%%
The induced effective action $S^{(2)}_{eff}$ is expected
to satisfy
%*** qed6 %%%%%%%%%%%%%%%%%%%%
\begin{eqnarray}
\frac{\del S^{(2)}_{eff}}{\del F^\mn}\equiv
<J_\mn>\sim e^2g_Y\ep_{\mn\ls\tau}\pl^\la\Phi 
{\bar \psihat}\Si^{\si\tau}\psihat
\pr
\label{qed6}
\end{eqnarray}
%%%%%%%%%%%%%%%%%%%%%%%%%%%%%
Then $S^{(2)}_{eff}$ is obtained as, in the thin wall limit, 
%*** qed7 %%%%%%%%%%%%%%%%%%%%
\begin{eqnarray}
S^{(2)}_{eff}
\sim e^2g_Y\ep_{\mn\ls\tau}\int d^5X\pl^\la\Phi\,F^\mn
{\bar \psihat}\Si^{\si\tau}\psihat\nn
= e^2g_Y\vz\ep_{\ab\gd}\int d^4x F^\ab
{\bar \psi}\si^{\gd}\psi
= -ie^2g_Y\vz\int d^4x F^\ab
{\bar \psi}\gago\si_{\ab}\psi
\pr
\label{qed7}
\end{eqnarray}
%%%%%%%%%%%%%%%%%%%%%%%%%%%%%
%Magnetic moment term ???? 
This term is the EDM term of
(\ref{ferKK6}). It violates the CP-symmetry. The coupling
depends on the vacuum expectation value of $\Phi$, 
$\vz=<\Phi>$ which are, at present, not known.
In order to estimate the magnitude of the
coupling, it is necessary to apply this model
to the quark-lepton (electro-weak) theory and fix
the value. We expect the magnitude could be
sufficiently large so that the result can be tested
by present or near-future experiments.

The appearance of the EDM term corresponds
to the CP-extremely-violated case (ii) $|\mhat|\gg \m$ of Sec.3. 
We compare the parameter relation with the thin wall relation
(\ref{yuka4b}) $k\gg 1/r_c$. 
Because the parameter $\m$ in KK case corresponds to $1/r_c$ in RS case,
we notice the 5D fermion mass in KK case ($|\mhat|$) corresponds to
the inverse of the thickness in RS case ($k$). 

The more fascinating view on the correspondence
is that 
{\it the RS approach and the KK one are "dual" each other}. 
We compare the above thin wall limit
with the small radius limit (i) $|\mhat|\ll \m$ of Sec.3. 
The thin wall limit, which is regarded as the dimensional reduction,
can be consistently taken in RS model and EDM term naturally appears
there. CP is {\it not preserved}. 
The theoretical treatment is justified as far as the relations
(\ref{yuka4b}) and (\ref{yuka4c}) are satisfied:\ 
$1/\sqrt[3]{G_5}\gg k\gg 1/r_c$. 
While, in the KK case, the dimensional reduction takes place
in the case (i) of Sec.3, 
which can be controlled theoretically as far as
the relation (\ref{ferKK8b})
$1/\sqrt[3]{G_5}=\sqrt[3]{100}\m\geq\m\gg|\mhat|$
is satisfied. MDM term
naturally appears and CP is {\it preserved} here.

In the effective action evaluation, the bulk Higgs field
$\Phi(y)$ plays an important role. It serves as a 
bridge between the bulk world and the 4D world.
%%%%%%%%%%%%%%%%%%%%%%%%%%%%%%%%%%%%%%%%%%%%%%%%%%%%%%%%%%%%%%%%%%%%%
%%%%      SEC. 8  Conclusion                       %%%%%%%%%%%%%%%%%%
%%%%%%%%%%%%%%%%%%%%%%%%%%%%%%%%%%%%%%%%%%%%%%%%%%%%%%%%%%%%%%%%%%%%%
\section{Discussion and Conclusion}
Recently the higher dimensional (5D,6D) theories
(with boundary terms) 
have been vigorously investigated in order to
make them some consistent {\it quantum} field theories.
In the past they were not so seriously examined
because the renormalizability condition strictly
excludes the higher dimensional theories.
(One of few exceptions is the success of 
the vacuum energy calculation of the 5D Kaluza-Klein theory
\cite{AC83PRD}.) 
This past attitude was changed by the progress
of the string and D-brane physics. Especially
the AdS/CFT correspondence\cite{Malda97} gave us
the image of the 4D world as a boundary of a higher
dimensional space-time. 
The most crucial point is to find how the ordinary 4D quantum
field theory is generalized to the extra-space direction. 
The treatment of the extra axis(es)
becomes the central concern. In this point, some interesting
proposals begin to be given by Goldberger-Wise\cite{GW0104} 
and by Randall-Schwartz\cite{RS0108a,RS0108b}.
Both regard the extra space as the scale parameter space
of the renormalization group. 
Renormalization flow moves in the extra space.
New regularizations\cite{Nibbe0108} are also proposed
in order to regularize the summation over the all
KK modes in the extra space direction.
We comment that, in connection with the chiral determinant
calculation 
based on the domain wall configuration in the 5D space-time, 
a new regularization is
proposed\cite{SI9908,SI9911}, where the extra space is regarded
as the temperature (proper time) space. 
The present approach to this bulk quantum theory is
different from the above ones. We take
the {\it induced effective action} method. It has been
used in relation to  
the chiral and Weyl anomalies.
Famous successful ones are 2D WZNW model derived
from the 2D QCD \cite{PW83PLB}
and 2D induced quantum gravity\cite{Pol87MPL}.
We have examined mainly the thin wall limit.
In order to examine the configuration off the limit,
we must take into account the new proposals mentioned above.
We should pursue this line of research.

We have comparatively examined the KK model and the RS model.
Both have attractive features as the higher dimensional
unification models. 
Using the Cartan formalism, fermions are properly treated.
The KK expansion and the RS expansion are compared.
%The comparison becomes useful when we reveal their
%characteristic points. 
The periodic functions appear in the former case, 
while the Bessel functions characteristically appear
in the latter case. The dual property is controlled by
two scale parameters, $\mhat$ and $\mu$ in the KK case
whereas $k$ and $r_c$ in the RS case.
In particular we stress that, as was pointed out by 
Thirring for the KK case, the CP-violation term naturally
appears also in the RS model.

Finally we list the correspondence in Table 1.
\nl
\nl

%%%%%%%%%%%%%%%%%%%%%%%%%%%%%%%%%%%%%%%%%%%%%%%%%%%%%%%%%%%%%%%%%%%%%%%%%%
%%%%%%%%%%%%%%%%%%%%%  Table 1  %%%%%%%%%%%%%%%%%%%%%%%%%%%%%
%%%%%%%%%%%%%%%%%%%%%%%%%%%%%%%%%%%%%%%%%%%%%%%%%%%%%%%%%%%%%%%%%%%%%%%%%%
\begin{tabular}{|c|c|c|}
\hline
             & Kaluza-Klein           & Randall-Sundrum               \\
\hline
electric charge  & $e=f\m$,\ $f$:free para. & $e$:free para. \\
\hline
U(1) sym.  & $y\ra y+\La(x),$ transl. in \{y\}& 
                       $A_m\ra A_m+\pl_m\La,$ internal sym. \\
           & $A_a(x)\ra A_a(x)+\frac{1}{f}\pl_a\La$ & $(X^m)=(x^a,y)$:\ fixed\\
\hline
asym. geometry  & $S^1\times {\cal M}_4$ & AdS$_5$\ $(\om,\mtil=g_Y\vz)$ 
%$\sim S^1\times {\cal M}_4$???? 
                                                    \\
\hline
           & a massive KK mode for  & 5D bulk Higgs vacuum      \\
           & charged fermion $\psi$,  &   $<\Phi>=\pm\vz, y\ra\pm\infty$ \\
vacuum     & 0-th KK modes for  &  kink sol., $Z_2$-symmetry       \\
           & $g_{ab}, A_a, \si$ and &           \\
           & neutral fermion          &             \\
\hline
4D fermion mass & $\sqrt{\mhat^2+\m^2}$ & $g_Y\vz\times$overlap-int.\\
\hline
physical scale & $\mhat$\ :5D fermion mass  & $k$\ :(thickness)$^{-1}$ \\
\hline
global size  & $\m^{-1}$\ :radius of extra $S^1$    &  $r_c$\ :IR cutoff   \\
             & $y\ra y+2\pi\m^{-1}$ , periodic & $-r_c\leq y\leq r_c$      \\
\hline
dimensional       & $\m\gg\mhat$        &  $k\gg 1/r_c$     \\
reduction cond.   & small radius limit  & thin-wall limit   \\
\hline
5D classical      & $1/\sqrt[3]{G_5}\sim \sqrt[3]{100}\m\gg\mhat$ 
                                        & $1/\sqrt[3]{G_5}\gg k$  \\
condition         &                     &   \\
\hline
mode functions   & $\e^{ik\m y},k\in {\bf Z}$ 
                               & $J_\n(m_kz),N_\n(m_kz),\n=|\half-\mtil/\om|$\\
in extra space   & periodic func. 
                             & $z\om=\e^{\om y}$, $k\in {\bf Z}$,Bessel func.\\
\hline
CP property      & MDM in small radius   & EDM in thin wall   \\
                 & limit, CP-preserved   & limit, CP-violated  \\
\hline
\multicolumn{3}{c}{\q}                                                 \\
\multicolumn{3}{c}{Table 1\ \ Comparison of 
KK model and RS model.  }\\
%\multicolumn{3}{c}{        }\\
\end{tabular}
%%%%%%%%%%%%%%%%%%%%%%%%%  END  of  Table 1 %%%%%%%%%%%%%%%%%%%%%%%%%%%%%

%\vspace{2cm}

\vs 1
%%%%%%%%%%%%%%%%%%%%%%%%%%%%%%%%%%%%%%%%%%%%%%%%%%%%%%%%%%%%%%
%%%%%%%%%%%%%%%%%%%%%  Acknowledgment %%%%%%%%%%%%%%%%%%%%%%%
%%%%%%%%%%%%%%%%%%%%%%%%%%%%%%%%%%%%%%%%%%%%%%%%%%%%%%%%%%%%%%
\begin{flushleft}
{\bf Acknowledgment}
\end{flushleft}
The author thanks K. Akama 
for some comments and precious information about the brane world.
He also thanks S.D. Odintsov for bringing the references 
\cite{RS0108a,RS0108b} to his attention. 
Parts of the present results were presented
at Chubu Summer School (Syuzenji, Japan, 2001.8.30-9.2),
the 5th KEK Topical Conference -Frontiers in Flavor Physics-
(Tsukuba, Japan, 2001.11.20-22),
YITP Workshop on "Fundamental Problems and Applications
in Quantum Field Theory" (Kyoto, Japan, 2001.12.19-21), 
the annual meeting of the Physical Society of Japan
(Ritsumeikan Univ.,Kusatsu, Japan, 2002.3.24-27) and
Summer Institute 2002 (Fuji-Yoshida, Japan, 2002.8.13-20). 
Discussions with the audience
are appreciated. In particular the author thanks 
T. Inami, T. Kugo, M. Nakahara and N. Nakanishi.

\vs 1
%%%%%%%%%%%%%%%%%%%%%%%%%%%%%%%%%%%%%%%%%%%%%%%%%%%%%%%%%%%%%%%%%%%%%
%%%%      App.A  5D Dirac Matrix                       %%%%%%%%%%%%%%%%%%
%%%%%%%%%%%%%%%%%%%%%%%%%%%%%%%%%%%%%%%%%%%%%%%%%%%%%%%%%%%%%%%%%%%%%
\section{Appendix A\ \ 5D Dirac Matrix}
In order to clearly discuss (5D) Dirac equation, we must
fix the representation of the gamma matrices.
The signs and phases appearing in the text depend
on the convention. 
We take the Majorana representation for the Dirac matrix.
%*** Dir1 %%%%%%%%%%%%%%%%%%%%
\begin{eqnarray}
\{\ga_\m ,\ga_\n\}=2\eta_\mn\com\q
(\eta_\mn)=\mbox{diag}(-1,1,1,1,1)\com\nn
\q\q\q\q\q \m,\n=0,1,2,3,5\nn
\ga_0=-\ga^0=\left(
\begin{array}{cc}
0       & i\si_2 \\
i\si_2  & 0
\end{array}
            \right)\com\q
\ga_1=\ga^1=\left(
\begin{array}{cc}
\si_1 & 0 \\
0     & \si_1
\end{array}
            \right)\com\nn
\ga_2=\ga^2=\left(
\begin{array}{cc}
0       & -i\si_2 \\
i\si_2  & 0
\end{array}
            \right)\com\q
\ga_3=\ga^3=\left(
\begin{array}{cc}
\si_3 & 0 \\
0     & \si_3
\end{array}
            \right)\com\nn
\ga_5=\ga^5=i\ga_0\ga_1\ga_2\ga_3=-i\ga^0\ga^1\ga^2\ga^3=
      \left(
\begin{array}{cc}
\si_2 & 0 \\
0     & -\si_2
\end{array}
            \right)\com
\label{Dir1}
\end{eqnarray}
%%%%%%%%%%%%%%%%%%%%%%%%%%%%%
where $\si_i (i=1,2,3)$ are the Pauli matrices:\ 
$\si_1=(0,1 / 1,0), \si_2=(0,-i / i,0), \si_3=(1,0 / 0,-1)$. 
The above gamma matrices have the following properties.
\begin{itemize}
 \item
$\ga_0,\ga_1,\ga_2,\ga_3$ are real matrices, while
$\ga_5$ is pure imaginary.
 \item
$\ga_1,\ga_2,\ga_3,\ga_5$ are hermitian, while
$\ga_0$ is anti-hermitian.
$\ga_0{\ga_\m}^\dag\ga_0=\ga_\m$
 \item
$\ga_1,\ga_2,\ga_3$ are symmetric, while
$\ga_0,\ga_5$ are anti-symmetric.
 \item
${\ga_0}^2=-1, {\ga_1}^2={\ga_2}^2={\ga_3}^2={\ga_5}^2=1$
\end{itemize}

\vs 1
%%%%%%%%%%%%%%%%%%%%%%%%%%%%%%%%%%%%%%%%%%%%%%%%%%%%%%%%%%%%%%%%%%%%%
%%%%      App.B Randall-Sundrum Geometry in the Cartan Formalism  %%%
%%%%%%%%%%%%%%%%%%%%%%%%%%%%%%%%%%%%%%%%%%%%%%%%%%%%%%%%%%%%%%%%%%%%%
\section{Appendix B\ \ Randall-Sundrum Geometry in the Cartan Formalism}
We take the more general metric than (\ref{RS1}) which is
useful when the coordinate $y$ taken in the text is
transformed to another coordinates. 
%*** Car1 %%%%%%%%%%%%%%%%%%%%
\begin{eqnarray}
ds^2=\e^{-2\si(z)}\eta_{ab}dx^adx^b+\frac{4}{F(z)^2}dz^2\com\nn
%-\infty<y<+\infty\ ,\ 
-\infty<x^a<+\infty\ ,\ 
(\eta_{ab})=\mbox{diag}(-1,1,1,1)
\com
\label{Car1}
\end{eqnarray}
%%%%%%%%%%%%%%%%%%%%%%%%%%%%%
where $a,b=0,1,2,3$. 
The coordinate taken in the text is the case:
$F(z)=2$. If we denote this coordinate as $y$ as in the text, 
its coordinate region is $-\infty<y<+\infty$ which shows
non-compactness of the extra space. 
We can transform to some {\it compact} coordinates by the
following coordinate transformations
\footnote{
These coordinates are important to give some renormalization
group interpretation to the RS model\cite{SI0008}.
}
:
1) $z=\half\tanh y, -\half<z<\half, F(z)=(1-4z^2)$\ ;\ 
2) $y=\tan(\pi z), -\half<z<\half, F(z)=(2/\pi)\cos^2(\pi z)$.
We now start with $F(z)$ %of (\ref{Car1}) 
unspecified. 
Then the 5D metric is given as
%*** Car2 %%%%%%%%%%%%%%%%%%%%
\begin{eqnarray}
ds^2
=\gh_{mn}dX^mdX^n\com\q (X^m)=(x^a, z)\ ,\ \nn
(\gh_{mn})=\left(
\begin{array}{cc}
\eta_{ab}\e^{-2\si} & 0 \\
0  & \frac{4}{F(z)^2}
\end{array}
            \right)\com
\label{Car2}
\end{eqnarray}
%%%%%%%%%%%%%%%%%%%%%%%%%%%%%
where $m,n=0,1,2,3,5$. 
From these the basis of the cotangent manifold T$_p^*$M, 
$\th^\m$, is given as
%*** Car3 %%%%%%%%%%%%%%%%%%%%
\begin{eqnarray}
ds^2
=\th^\m\th^\n\etahat_\mn\ ,\ 
(\etahat_\mn)=\mbox{diag}(-1,1,1,1,1)\com\nn
\th^\al=\e^{-\si}\eta^\al_a dx^a\com\q \th^\fbar=\frac{2}{F(z)}dz
\com
\label{Car3}
\end{eqnarray}
%%%%%%%%%%%%%%%%%%%%%%%%%%%%%
where $\m,\n=0,1,2,3,\fbar$ and $\al=0,1,2,3$. 
From the above result we can read the f\"unf-bein $\e^\m_{~m}$.
%*** Car4 %%%%%%%%%%%%%%%%%%%%
\begin{eqnarray}
\th^\m=\e^\m_{~m}dX^m\com\nn
(\e^\m_{~m})=\left(
\begin{array}{cc}
\eta^\al_{~a}\e^{-\si} & 0 \\
0  & \frac{2}{F(z)}
\end{array}
            \right)\com\q
(\e_\m^{~m})=\left(
\begin{array}{cc}
\eta_\al^{~a}\e^{\si} & 0 \\
0  & \half F(z)
\end{array}
            \right)\com
\label{Car4}
\end{eqnarray}
%%%%%%%%%%%%%%%%%%%%%%%%%%%%%
where $\e_\m^{~m}$ is the inverse of $\e^\m_{~m}$.
The basis of the tangent manifold T$_p$M, $\e_\m$, is given by
%*** Car5 %%%%%%%%%%%%%%%%%%%%
\begin{eqnarray}
\e_\m\equiv \e_\m^{~m}\frac{\pl}{\pl X^m}\com\nn
\e_\al=\e^\si \eta_\al^{~a}\frac{\pl}{\pl x^a}\com\q
\e_\fbar=\half F(z)\frac{\pl}{\pl z}
\pr
\label{Car5}
\end{eqnarray}
%%%%%%%%%%%%%%%%%%%%%%%%%%%%%
$\{\e_\m\}=\{\e_\al,\e_\fbar\}$ and
$\{\th^\m\}=\{\th^\al,\th^\fbar\}$
constitute the non-coordinate basis, while
$\{\e_m\equiv \frac{\pl}{\pl X^m}\}
=\{\frac{\pl}{\pl x^a},\frac{\pl}{\pl z}\}$ and
$\{dX^m\}=\{dx^a,dz\}$ constitute the coordinate basis. 

The Lie algebra between $\{\e_\m\}$ is obtained as
%*** Car6 %%%%%%%%%%%%%%%%%%%%
\begin{eqnarray}
[\e_\al,\e_\be]=[\e_\fbar,\e_\fbar]=0,\ \ %\nn
[\e_\al, \e_\fbar ] = \half F\si'\e_\al
\com
\label{Car6}
\end{eqnarray}
%%%%%%%%%%%%%%%%%%%%%%%%%%%%%
where "$~'~$"=$\frac{d}{dz}$. 
The "structure constants", $c_\mn^{~~\la}$, are read as
%*** Car7 %%%%%%%%%%%%%%%%%%%%
\begin{eqnarray}
[\e_\m,\e_\n]=c_\mn^{~~\la}\e_\la ,\nn
c_\ab^{~~\m}=c_{\fbar\fbar}^{~~\m}=0\ ,\ 
c_{\al\fbar}^{~~\be}=-c_{\fbar\al}^{~~\be}
=\half F\si'\eta_\al^{~\be}\ ,\ 
c_{\al\fbar}^{~~\fbar}=-c_{\fbar\al}^{~~\fbar}=0
\pr
\label{Car7}
\end{eqnarray}
%%%%%%%%%%%%%%%%%%%%%%%%%%%%%
The first Cartan's structure equation, for the torsionless
case, is given by
%*** Car8 %%%%%%%%%%%%%%%%%%%%
\begin{eqnarray}
d\th^\m+\om^\m_{~\n}\wedge\th^\n=T^\m=0
\com
\label{Car8}
\end{eqnarray}
%%%%%%%%%%%%%%%%%%%%%%%%%%%%%
where $\om^\m_{~\n}$ is the {\it connection one-form}. Using
the obtained expression of $\th^\m$, the above
equation serves for fixing the connection.
%*** Car9 %%%%%%%%%%%%%%%%%%%%
\begin{eqnarray}
(i)\ \underline{\m=\al\ ,\ \th^\al=\e^{-\si}dx^\al}\q:\q
\om^\al_{~\fbar}=-\om_\fbar^{~\al}=-\frac{\si'}{2}F\th^\fbar\com\q
\om^\al_{~\be}=0\com\nn
(ii)\ \underline{\m=\fbar\ ,\ \th^\fbar=\frac{2}{F(z)}dz}\q :\q
\om^\fbar_{~\al}=\frac{\si'}{2}F\th_\al\com\q\th_\al=\eta_\ab\th^\be\nn
(iii)\ \underline{\om_\mn=-\om_{\n\m}\ ,\ \om^\m_{~\n}=-\om_\n^{~\m}}\q :\q
\om^\fbar_{~\fbar}=0
\pr
\label{Car9}
\end{eqnarray}
%%%%%%%%%%%%%%%%%%%%%%%%%%%%% 
Now we can calculate the curvature using the second
Cartan's structure equation.
%*** Car10 %%%%%%%%%%%%%%%%%%%%
\begin{eqnarray}
d\om^\m_{~\n}+\om^\m_{~\la}\wedge\om^\la_{~\n}=
R^\m_{~\n}=\half R^\m_{~\n\ls}\th^\la\wedge\th^\si
\com
\label{Car10}
\end{eqnarray}
%%%%%%%%%%%%%%%%%%%%%%%%%%%%%
where $R^\m_{~\n}$ is the curvature 2-form.
They are given as
%*** Car11 %%%%%%%%%%%%%%%%%%%%
\begin{eqnarray}
(i) \underline{\m=\al,\n=\be}\q :\nn
R^\al_{~\be\fbar\fbar}=R^\al_{~\be\ga\fbar}=-R^\al_{~\be\fbar\ga}=0\ ,\ 
R^\al_{~\be\ga\del}=-\frac{{\si'}^2}{4}F^2
(\eta^\al_\ga\eta_{\be\del}-\eta^\al_\del\eta_{\be\ga})\ ,\nn
(ii) \underline{\m=\fbar,\n=\fbar}\q :\q
R^\fbar_{~\fbar\mn}=0\com\nn
(iii) \underline{\m=\al,\n=\fbar}\q :\q
R^\al_{~\fbar\fbar\be}=-R^\al_{~\fbar\be\fbar}=
-\fourth (\si'F\e^{-\si})'F\e^\si \eta^\al_\be\com\nn
(iv) \underline{\m=\fbar\ ,\ \n=\al}\q :\q
R^\fbar_{~\al\fbar\be}=-R^\fbar_{~\al\be\fbar}=
\fourth (\si'F\e^{-\si})'F\e^\si \eta_\ab
\pr
\label{Car11}
\end{eqnarray}
%%%%%%%%%%%%%%%%%%%%%%%%%%%%%
The coordinate components are obtained by multiplying the
f\"unf-beins. 
%*** Car12 %%%%%%%%%%%%%%%%%%%%
\begin{eqnarray}
R^m_{~nkl}=\e_\m^{~m}\e^\n_{~n}\e^\la_{~k}\e^\si_{~l}R^\m_{~\n\ls}\com\q
R_{mn}=R^k_{~mnk}\com\nn
R_{55}=-\frac{4}{F}(\si''F+\si'F'-{\si'}^2F)\com\nn
R_{ab}=-\fourth F\e^{-2\si}(\si''F+\si'F'-4{\si'}^2F)\eta_{ab}\com\q
R_{5a}=R_{a5}=0
\pr
\label{Car12}
\end{eqnarray}
%%%%%%%%%%%%%%%%%%%%%%%%%%%%%
From the connection 1-form $\om^\m_{~\n}$, we can define
the connection coefficients as follows.
%*** Car13 %%%%%%%%%%%%%%%%%%%%
\begin{eqnarray}
\om^\m_{~\n}=\Ga^\m_{~\la\n}\th^\la\nn
(i)\ \underline{\om^\al_{~\fbar}=-\frac{\si'}{2}F\th^\al}\q :\q
\Ga^\al_{~\be\fbar}=-\frac{\si'}{2}F\eta^\al_\be\com\q
\Ga^\al_{~\fbar\fbar}=0\pr\nn
(ii)\ \underline{\om^\fbar_{~\fbar}=\om^\al_{~\be}=0}\q :\q
\Ga^\fbar_{~\m\fbar}=\Ga^\al_{~\m\be}=0\pr
\q (\mbox{especially}\ \ \Ga^\al_{~\fbar\be}=0\ .)\nn
(iii)\ \underline{\om^\fbar_{~\al}=\frac{\si'}{2}F\th_\al}\q :\q
\Ga^\fbar_{~\be\al}=\frac{\si'}{2}F\eta_{\be\al}\com\q
\Ga^\fbar_{~\fbar\al}=0
\pr
\label{Car13}
\end{eqnarray}
%%%%%%%%%%%%%%%%%%%%%%%%%%%%%
Note that generally $\Ga^\m_{~\la\n}\neq\Ga^\m_{~\n\la}$. 
Indeed the above result satisfies the torsionless condition,
for example, 
$T^\al_{~\fbar\be}\equiv \Ga^\al_{~\fbar\be}-\Ga^\al_{~\be\fbar}
-c_{\fbar\be}^{~~\al}=0$.

The connection coefficient $\Ga^\la_{~\mn}$ are
related with the same one with respect to $\{e_m\}$ as
%*** Car14 %%%%%%%%%%%%%%%%%%%%
\begin{eqnarray}
\Ga^\la_{~\mn}=\e^\la_{~n}\e_\m^{~m}(\pl_m\e_\n^{~n}+
\e_\n^{~k}\Ga^n_{~mk})\com\nn
\Ga^k_{~mn}=\e_\la^{~k}\e^\m_{~m}\e^\n_{~n}\Ga^\la_{~\mn}
-\pl_m\e_\m^{~k}\e^\m_{~n}
\pr
\label{Car14}
\end{eqnarray}
%%%%%%%%%%%%%%%%%%%%%%%%%%%%%
This formula gives
%*** Car15 %%%%%%%%%%%%%%%%%%%%
\begin{eqnarray}
\Ga^5_{~55}=-\frac{F'}{F}\com\q
\Ga^5_{~ab}=\frac{\si'}{4}F^2\e^{-2\si}\eta_{ab}\com\q
\Ga^a_{~5b}=\Ga^a_{~b5}%????
=-\si'\eta^a_{~b}
\pr
\label{Car15}
\end{eqnarray}
%%%%%%%%%%%%%%%%%%%%%%%%%%%%%
The connection coefficient with the "mixed type" $\Ga^\m_{~m\n}$,
is defined by 
%*** Car16 %%%%%%%%%%%%%%%%%%%%
\begin{eqnarray}
\Ga^\m_{~m\n}\equiv \Ga^\m_{~\la\n}\e^\la_{~m}\com\nn
\om^\m_{~\n}=\Ga^\m_{~\la\n}\th^\la=\Ga^\m_{~\la\n}\e^\la_{~m}dX^m
=\Ga^\m_{~m\n}dX^m
\pr
\label{Car16}
\end{eqnarray}
%%%%%%%%%%%%%%%%%%%%%%%%%%%%%
This formula gives
%*** Car17 %%%%%%%%%%%%%%%%%%%%
\begin{eqnarray}
(i)\ \underline{\om^\fbar_{~\fbar}=0\com\q\om^\al_{~\be}=0}\q :\q
\Ga^\fbar_{~m\fbar}=\Ga^\al_{~m\be}=0\com\nn
(ii)\ \underline{\om^\al_{~\fbar}=-\frac{\si'}{2}F\e^{-\si}dx^\al}\q :\q
\Ga^\al_{~5\fbar}=0\com\q 
\Ga^\al_{~a\fbar}=-\frac{\si'}{2}F\e^{-\si}\eta^\al_a\com\nn
(iii)\ \underline{\om^\fbar_{~\al}=\frac{\si'}{2}F\e^{-\si}dx^\al}\q :\q
\Ga^\fbar_{~5\al}=0\com\q 
\Ga^\fbar_{~a\al}=\frac{\si'}{2}F\e^{-\si}\eta_{a\al}
\pr
\label{Car17}
\end{eqnarray}
%%%%%%%%%%%%%%%%%%%%%%%%%%%%%

\vs 1
%%%%%%%%%%%%%%%%%%%%%%%%%%%%%%%%%%%%%%%%%%%%%%%%%%%%%%%%%%%%%%%%%%
%%%%%%%%%%%%%%%%%%%%%%%% reference %%%%%%%%%%%%%%%%%%%%%%%%%%%%%%%
%%%%%%%%%%%%%%%%%%%%%%%%%%%%%%%%%%%%%%%%%%%%%%%%%%%%%%%%%%%%%%%%%%

\end{document}